\RequirePackage{lineno}
\documentclass[aps,prl,showpacs,amsfonts,notitlepage,amssymb,amsmath,nofootinbib,superscriptaddress,longbibliography,twocolumn]{revtex4-2}
\usepackage[utf8]{inputenc}
\usepackage[english]{babel}
\usepackage[titletoc,toc,title]{appendix}
\usepackage{amsmath}
\usepackage{amsfonts}
\usepackage{amssymb}
\usepackage[colorlinks=true,linkcolor=blue,citecolor=blue,urlcolor=blue]{hyperref}
\usepackage{graphicx}
\usepackage{float}
\usepackage{enumerate}
\usepackage{csquotes}
\usepackage[caption=false]{subfig}
\usepackage{dsfont}
\usepackage{color}
\usepackage[dvipsnames]{xcolor}
\usepackage{bbold}
\usepackage{mathtools}
\usepackage{tensor}
\usepackage[export]{adjustbox}
\usepackage{hyperref}
\usepackage{tikz}
\usepackage{comment}
\usepackage{colortbl}
\DeclareMathOperator{\Arcsin}{Arcsin}
\DeclareMathOperator{\Arccos}{Arccos}
\makeatletter

    \def\CT@@do@color{%
      \global\let\CT@do@color\relax
            \@tempdima\wd\z@
            \advance\@tempdima\@tempdimb
            \advance\@tempdima\@tempdimc
    \advance\@tempdimb\tabcolsep
    \advance\@tempdimc\tabcolsep
    \advance\@tempdima2\tabcolsep
            \kern-\@tempdimb
            \leaders\vrule
                    \hskip\@tempdima\@plus  1fill
            \kern-\@tempdimc
            \hskip-\wd\z@ \@plus -1fill }
    \makeatother
\begin{document}

\title{How to create analogue black hole or white fountain horizons and LASER cavities\\ in experimental free surface hydrodynamics?}
\date{\today}

\author{Alexis Bossard}
\affiliation{Institut Pprime, UPR 3346 CNRS--Universit\'{e} de Poitiers--ISAE ENSMA, 11 Boulevard Marie et Pierre Curie--T\'{e}l\'{e}port 2--BP 30179, 86962 Futuroscope Chasseneuil Cedex, France}
\author{Nicolas James}
\affiliation{Laboratoire de Math\'{e}matiques et Applications (UMR 7348), CNRS--Universit\'{e} de Poitiers, 11 Boulevard Marie et Pierre Curie--T\'{e}l\'{e}port 2--BP 30179, 86962 Futuroscope Chasseneuil Cedex, France}
\author{Camille Aucouturier}
\affiliation{Institut Pprime, UPR 3346 CNRS--Universit\'{e} de Poitiers--ISAE ENSMA, 11 Boulevard Marie et Pierre Curie--T\'{e}l\'{e}port 2--BP 30179, 86962 Futuroscope Chasseneuil Cedex, France}
\author{Johan~Fourdrinoy}
\affiliation{Institut Pprime, UPR 3346 CNRS--Universit\'{e} de Poitiers--ISAE ENSMA, 11 Boulevard Marie et Pierre Curie--T\'{e}l\'{e}port 2--BP 30179, 86962 Futuroscope Chasseneuil Cedex, France}
\author{Scott Robertson}
\affiliation{Institut Pprime, UPR 3346 CNRS--Universit\'{e} de Poitiers--ISAE ENSMA, 11 Boulevard Marie et Pierre Curie--T\'{e}l\'{e}port 2--BP 30179, 86962 Futuroscope Chasseneuil Cedex, France}
\author{Germain Rousseaux}
\affiliation{Institut Pprime, UPR 3346 CNRS--Universit\'{e} de Poitiers--ISAE ENSMA, 11 Boulevard Marie et Pierre Curie--T\'{e}l\'{e}port 2--BP 30179, 86962 Futuroscope Chasseneuil Cedex, France}

\begin{abstract}

Transcritical flows 
in free surface hydrodynamics emulate black hole horizons and their time-reversed versions known as white fountains. Both analogue horizons have been shown to emit 
Hawking radiation, the amplification of 
waves via scattering 
at the horizon. Here we report on an experimental validation of the hydrodynamic laws that govern transcritical flows, for the first time in a free surface water channel using an analogue space-time geometry controlled by a bottom obstacle. A prospective study, both experimental and numerical, with a second obstacle downstream of a first one is presented to test in the near-future the analogous black hole laser instability, namely the super-amplification of Hawking radiation by successive bounces on a pair of black and white horizons within cavities which allow the presence of negative energy modes necessary for the amplification process. Candidate hydrodynamic regimes are discussed thanks to a phase diagram based on the scaled relative heights of both obstacles and the ratio of flow to wave speed in the upstream region.
\end{abstract}

\maketitle
William Unruh's demonstration in 1981 \cite{Unruh81} of a mathematical analogy between the propagation of light in a curved space-time and the propagation of acoustic waves in a moving fluid has led to the implementation of many experiments, not just in the original domain of acoustics. Indeed, the analogy, called Analogue Gravity \cite{LivingReview, barcelo2019analogue}, can be extended to other fields of physics such as optics \cite{aguero2020hawking}, Bose-Einstein condensates \cite{garay2000sonic} and interfacial hydrodynamics \cite{rousseaux2020classical}. In the latter, surface waves play the role of light and an accelerating fluid plays the role of an effective black hole space-time \cite{schutzhold2002gravity}. In the case of a Schwarzschild black hole in general relativity, the horizon is defined as the position at which the escape velocity 
($v_\text{escape}(r)=\sqrt{2GM/r}$ where $M$ is the mass of the black hole and $G$ is Newton's 
gravitational constant) is equal to the speed of light. 
An analogue definition of a hydrodynamical horizon is used in interfacial hydrodynamics based on the Froude number $Fr=U/\sqrt{gh}$ which reaches unity when the velocity of long waves, $c(x)=\sqrt{gh(x)}$, is equal to the velocity of the flow current, $U(x)$. The flow is then said to be {\it transcritical}. In the case where the Froude number is less (respectively greater) than 1, the flow is characterized as {\it subcritical} (respectively {\it supercritical}). Since waves at the surface of a fluid are in general dispersive, the notion of an analogous dispersionless horizon like in General Relativity is generalized with the introduction of a dispersive group velocity horizon \cite{nardin2009wave, rousseaux2010horizon, rousseaux2020classical}, defined as $Fr_\text{disp}=U/v_g$, where $v_g=\partial \left(\omega-\overrightarrow{U}\cdot\overrightarrow{k}\right)/\partial k$ is the group velocity in the rest frame of the flow. 

With this kinematical analogy between several fields of physics, Analogue Gravity allows us to verify some theoretical predictions, like the Hawking radiation of black holes 
\cite{Hawking}. 
This is the predicted glow that originates in quantum vacuum fluctuations 
in the vicinity of the event horizon, with the partners of each virtual pair being separated by the horizon so that real particles are emitted. 
Spontaneous Hawking radiation, seeded by quantum noise in the manner just described, has been measured in a Bose-Einstein condensate with a dispersive superluminal linear and transcritical regime \cite{steinhauer2016observation, Steinhauer-2019}. 
However, a classical version of it has been measured with hydrodynamic analogues in the dispersive 
subcritical regime, via either the amplification of turbulent noise~\cite{euve2016observation}, or by stimulating waves in interfacial hydrodynamics experiments at a white fountain horizon \cite{rousseaux2020classical}.

Another theoretical prediction, also relying on Hawking radiation, is the black hole LASER effect or instability.  This is the amplification or damping of Hawking radiation due to successive bouncing of trapped modes between two horizons, which would act as active mirrors in a LASER cavity. One of the first theoretical contributions on the LASER instability due to Vilenkin is the coupling between the event horizon of a rotating black hole and a convex mirror, in order to amplify gravitational, scalar or electromagnetic waves \cite{Vilenkin}. In the case of the more recent Corley \& Jacobson scenario \cite{CorleyJacobson}, the convex mirror is replaced by an inner horizon (which happens, for instance, with charged black holes).  In this case, the partner particles, emitted on the supercritical side of the outer horizon, reach the inner horizon and are then reflected, going on to reach the outer horizon again and further stimulate the Hawking radiation process at each reflection. This stimulation depends on the nature of the particles: amplification occurs for bosons and damping for fermions. Moreover, in order to be able to return to the outer horizon, the partner particles in the supercritical region need a superluminal dispersive correction in order to beat the 
dragging effect of space-time within the outer horizon. The possibility of a black hole lasing effect in an analogous system has been demonstrated numerically in the case of nonlinear optics \cite{Faccio}, in interfacial hydrodynamics \cite{Peloquin}, and in Bose-Einstein condensates \cite{Leonhardt}. In order to accelerate a flow and thus create an analogous black hole, a geometrical streamlined obstacle is used that modifies the bathymetry. Hence, a study with two obstacles is necessary to create LASER cavities delimited by successive time-reversed horizons, generated by the acceleration or deceleration induced by the obstacles.  

The remainder of the article is thus organised accordingly: in the one-obstacle case, it is shown that a black hole type flow is formed, without imposed downstream boundary condition and without initial static water depth. We show experimentally that the flow respects the so-called theoretical Long's law relating flow physics to geometry, a premiere in the literature. In the two-obstacle case, a phase diagram is constructed and modernised, based on an existing diagram introduced previously in a pure hydrodynamics context, and new hydrodynamic regimes and candidate regimes for the analogous black hole LASER effect are demonstrated for the first time.

{\it Flows over one obstacle.}
In order to create a hydrodynamic horizon in an open water channel, the acceleration or deceleration of the flow must be forced. A simple method is to change the bathymetry of the channel by placing an obstacle on the bottom. The study of a flow over variable bathymetry is not a simple problem and has a long history~\cite{Lowery-Liapis-1999, rousseaux2020classical, Binder, Baines}. To simplify the problem, the current is assumed to be irrotational, stationary and the longitudinal velocity $U$ is related to the mean water depth $h$ by the conservation of flow rate (assuming a laminar plug-like velocity profile in the vertical direction) $U(x)=\frac{Q}{Wh(x)}$ where $W$ is the width of the rectangular open water channel, $Q$ is the flow rate produced by the pump and $x$ is the position along the channel. Moreover, to obtain an analogue dispersionless black hole horizon, the local Froude number must be equal to 1, i.e. a so-called transcritical regime is required. In hydraulics, Long's 1954 formula is the simplest theory that allows the separation of transcritical, subcritical and supercritical regimes as a function of the so-called obstruction factor~\cite{Long, Baines} $r=b_\text{max}/h_\text{upstream}$, with $b_\text{max}$ being the maximum height of the obstacle and where $h_\text{upstream}$ is the upstream water depth before the bottom obstacle taken to be constant at a given distance of the obstruction. In practice, the upstream water depth changes because of friction: turbulent head losses, bottom and side frictions due to viscosity along the water channel. Long's formula assumes no viscosity, flow rate conservation and energy conservation~\cite{Long, Baines}. The upstream Froude number is equal to $Fr_\text{up}=U_\text{upstream}/\sqrt{gh_\text{upstream}}$ and a transcritical regime is obtained ($Fr(x)=1$ somewhere on the obstacle) when~\cite{Long}:
\begin{linenomath}
\begin{equation}
r_\text{Long}=1+\frac{1}{2}Fr_\text{up}^2-\frac{3}{2}Fr_\text{up}^{\frac{2}{3}}
\label{Long}
\end{equation}
\end{linenomath}
Explicit formulas $Fr_\text{up}=f(r)$ can be obtained by looking for the roots of the implicit equation~\ref{Long}. These are the 1948 Schijf's expressions~\cite{schijf1949protection} obtained before Long in a different context namely boat navigation in confined waterways and that we relate for the first time in the literature (a bottom obstacle is similar to an inverted ship occupying the full width of a rectangular water channel):

\begin{linenomath}
\begin{align*}
    Fr_\text{up}=\left(2\sin\left(\frac{\pi-\Arcsin(1-r_\text{Long})}{3}\right)\right)^{\frac{3}{2}}\,,\text{for}\quad Fr_\text{up}\geqslant1  \\ 
    Fr_\text{up}=\left(2\sin\left(\frac{\Arcsin(1-r_\text{Long})}{3}\right)\right)^{\frac{3}{2}}\,,\text{for}\quad Fr_\text{up}\leqslant1
\label{Schijf}
\end{align*}
\end{linenomath}
The upper/lower formula separates the transcritical and the supercritical/subcritical regimes. Only the latter is relevant for Analogue Gravity to generate transcritical flows akin to hydraulic black hole as we will demonstrate experimentally below. We stress that the flow is turbulent since the Reynolds number is large. We will show that Long's laminar flow formula still applies quite well to the mean speed of the turbulent current and fluctuations are supposed to be small because of the streamlined bottom obstacle.

An approximation of the Long's 1954 formula can be constructed when the curvature of the obstacle is small compared to the curvature of the flow ($1/R \ll 1/h$) and the upstream velocity is small compared to the velocity above the obstacle, then the Faber's 1995 approximation~\cite{Faber} is obtained ($r\rightarrow 1$):
\begin{linenomath}
\begin{equation}
    h_\text{upstream}=b_\text{max}+\frac{3}{2}\sqrt[3]{\frac{q^2}{g}}
    \label{Faber}
\end{equation}
\end{linenomath}
To obtain the scaled form, we divided the equation~\ref{Faber} by the upstream height, $h_\text{upstream}$:
\begin{linenomath}
\begin{equation}
    r_\text{Faber}=1-\frac{3}{2}Fr_\text{up}^{\frac{2}{3}}
    \label{Faber_scale}
\end{equation}
\end{linenomath}
that lacks the square term compared to Long's formula.

Other theories have also been discussed theoretically, based on the analysis of the forced KdV (fKDV) equation, which takes into account dispersive effects such as in the work of Keeler and Binder \cite{keeler, Binder}. For instance, Binder deduced another explicit formula relating the Froude number with the obstruction ratio:

\begin{linenomath}
\begin{equation}
    Fr_\text{fKDV}=1-\left(\frac{9r_\text{fKDV}}{4\sqrt{2}}\right)^{\frac{2}{3}}
    \label{KDV}
\end{equation}
\end{linenomath}

To compare, experimentally, these theories predicting the onset of the transcritical regime or a hydrodynamics black hole horizon, several obstacles used in anterior studies on Analogue Gravity were employed (see the Supplemental Material). They are placed within the channel, without imposing a downstream boundary condition or a static height contrarily to most of previous setups reported so far \cite{Rousseaux-et-al-2008, rousseaux2010horizon, Weinfurtner-et-al-2011, Euve-et-al-2015, euve2016observation, Euve-Rousseaux-2017, euve2020scattering} except \cite{fourdrinoy2022correlations}. The steady upstream height is measured for a given obstacle with a graduated ruler (with uncertainty $\pm 1mm$) as a function of an increasing flow rate. This procedure is repeated for different geometries and obstacles sizes. We plot then $Fr_\text{up}(r)$.

\begin{figure*}
\includegraphics[scale=0.7,width=0.95\textwidth]{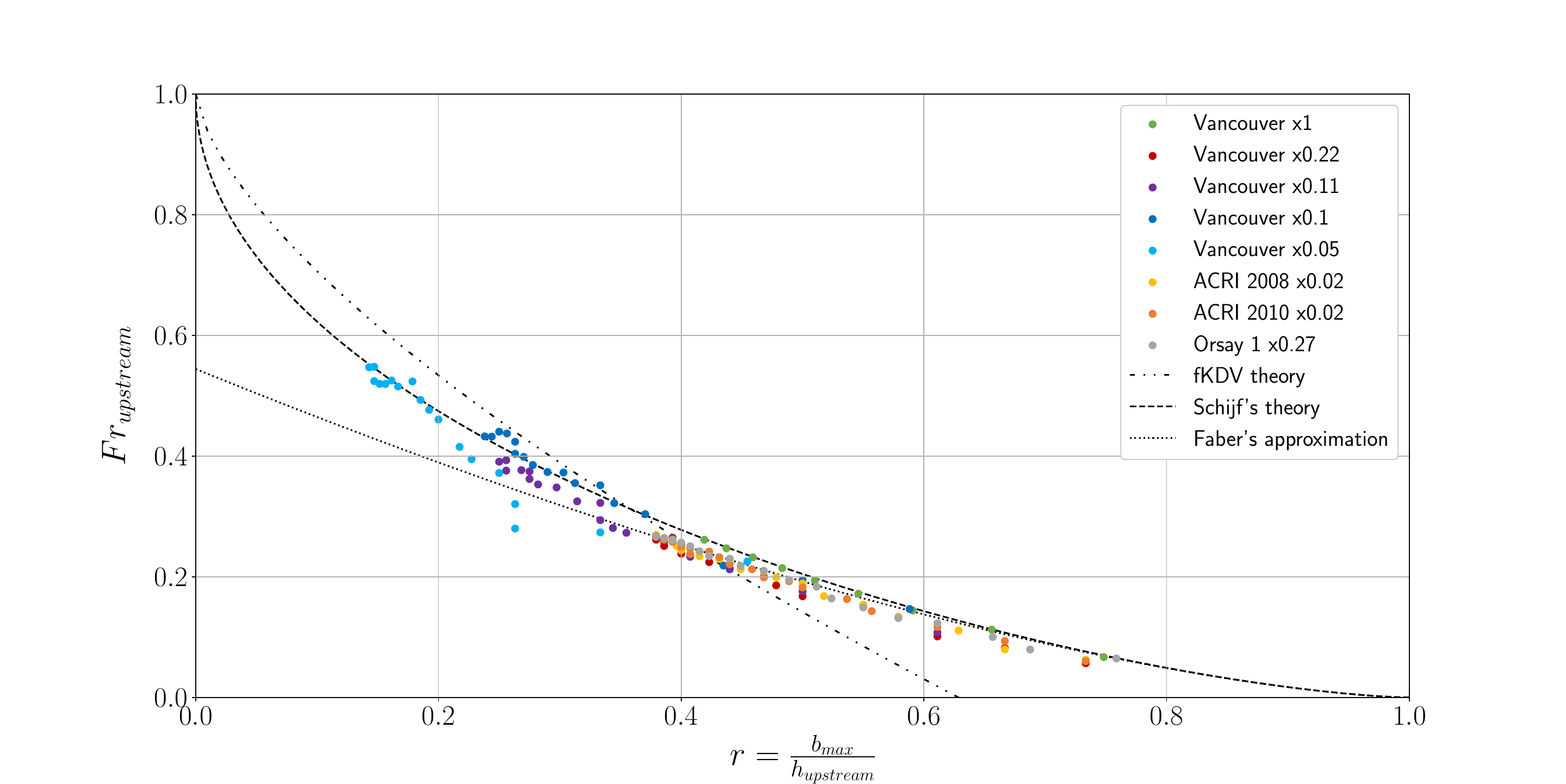}
\caption{Experimental diagram of the upstream Froude number as a function of the obstruction ratio $\textup{Fr}_\text{up}(r)$ obtained in a free surface channel, without downstream condition, without static water height and with asymmetric obstacles. The experimental points are the coloured points and each colour corresponds to obstacles of different sizes and shapes, which are compared to the Long's theory (identical to the Schijf's theory), the Binder's theory and the Faber's approximation.
\label{fig:camille}}
\end{figure*}

The experimental results are shown in the diagram \ref{fig:camille} superposed to theoretical models. As expected, Faber's approximation tends to Long's theory when $Fr_{up}\rightarrow 0$ (or $r\rightarrow 1$). The diagram gathers the results for different geometries and sizes (extra graphics separating the influence of geometry and size are given in the Supplemental Material). Even if there are differences in the geometries, the experimental points are close to the limits predicted by Long and its Faber approximation, which means that these points are transcritical flows as seen by visual inspection. The points closest to the Long's theory are the green points for the bigger obstacle, Vancouver $1$, in the bigger channels used previously by \cite{Weinfurtner-et-al-2011, Euve-et-al-2015, euve2016observation, Euve-Rousseaux-2017, euve2020scattering}. The points for the smaller obstacles in the smaller channel used in this work and previously by \cite{fourdrinoy2022correlations} are further away from Long's theoretical curve, because the boundary layer for these obstacles is more developed and therefore the plug-like velocity profile assumed for Long's law is no longer completely valid. The maximum height of the obstacles is also an important parameter: the smaller the obstacle, the larger the range of values of the Froude number explored. This is due to the fact that a larger obstacle generates a larger return flow and therefore for a slower flow a larger dynamic height upstream can be reached. An example of the consequence of the maximum obstacle height is illustrated with for instance the Vancouver $0.05$. This obstacle has the lowest height and the range explored is $r\in\mathclose{[}0.14\,;0.46\mathclose{]}$ and $Fr_\text{up}\in\mathclose{[}0.23\,;0.55\mathclose{]}$. Conversely, for the Vancouver $1$, the range explored is $r\in\mathclose{[}0.42\,;0.75\mathclose{]}$ and $Fr_\text{up}\in\mathclose{[}0.07\,;0.26\mathclose{]}$. The Orsay and ACRI more streamlined geometries are closer to theoretical prediction than the Vancouver geometry. The bigger obstacles are better when compared to Long's theory than smaller obstacles because of the smaller value of the ratio between the boundary layer and the total depth for the former.

{\it Flows over two obstacles.}
The study of flows over two successive obstacles relies on the seminal work of Pratt~\cite{pratt1982dynamics, Pratt1984}. Indeed, he looked to the influence of a second obstacle downstream of the first. He fixed the position and shape of the first obstacle. The distance between both obstacles was four times the length of the first one. As for the second obstacle, the length and shape were variable thanks to a tongue at the back of the obstacle, which allowed him, when the tongue was pulled, to modify the shape and the maximum height of the obstacle. The procedure of his experiments was to fix the flow rate, the first obstacle and the inter-obstacle distance and to explore the different hydrodynamic regimes by changing the length of the second obstacle.  Once this procedure is applied, he built a diagram ($Fr_\text{up}$ as a function of $\Delta b=(b_2-b_1)/b_1$), where $b_i$ is the maximum height of the $\text{i}^\text{th}$ obstacle. We introduce in the literature, the now called Pratt number terminology $\mathcal{P}=(b_2-b_1)/b_1$ that replaces the Pratt notation $\Delta b$ used in 1984~\cite{Pratt1984}.

To modernise his diagram, we reproduce Pratt's experiments within an Analogue Gravity context probing a larger range of parameters being more accurate. For this purpose, the same free-surface channel is used as for the single-obstacle experiments. However, the geometry of the obstacle is fixed to the one in \cite{rousseaux2010horizon}, ACRI 2010, a most simple and reproducible obstacle with three slopes and left-right asymmetry to avoid flow recirculation. A homothetic ratio is used on all lengths of the geometry (except the width $W$) to produce smaller obstacles and therefore a smaller maximum height to probe a bigger range of Pratt's number to be scanned, $\mathcal{P}\in\mathclose{[}-1\,;1\mathclose{]}$ . In doing so, the first and second obstacle dimensions as well as the flow rate are varied contrary to Pratt's experiment. To construct a  modernized diagram, both obstacles are placed in the channel at an arbitrarily chosen distance of 9.2 cm. This distance is measured from the end of the first obstacle to the beginning of the second and is fixed for all experiments. The points in the diagram, $Fr_\text{up}(\mathcal{P})$, have been probed by strictly increasing the flow rate and waiting for the steady state to avoid any hysteresis phenomena.

\begin{turnpage}
\begin{figure*}
\includegraphics[scale=0.52,width=1.35\textwidth]{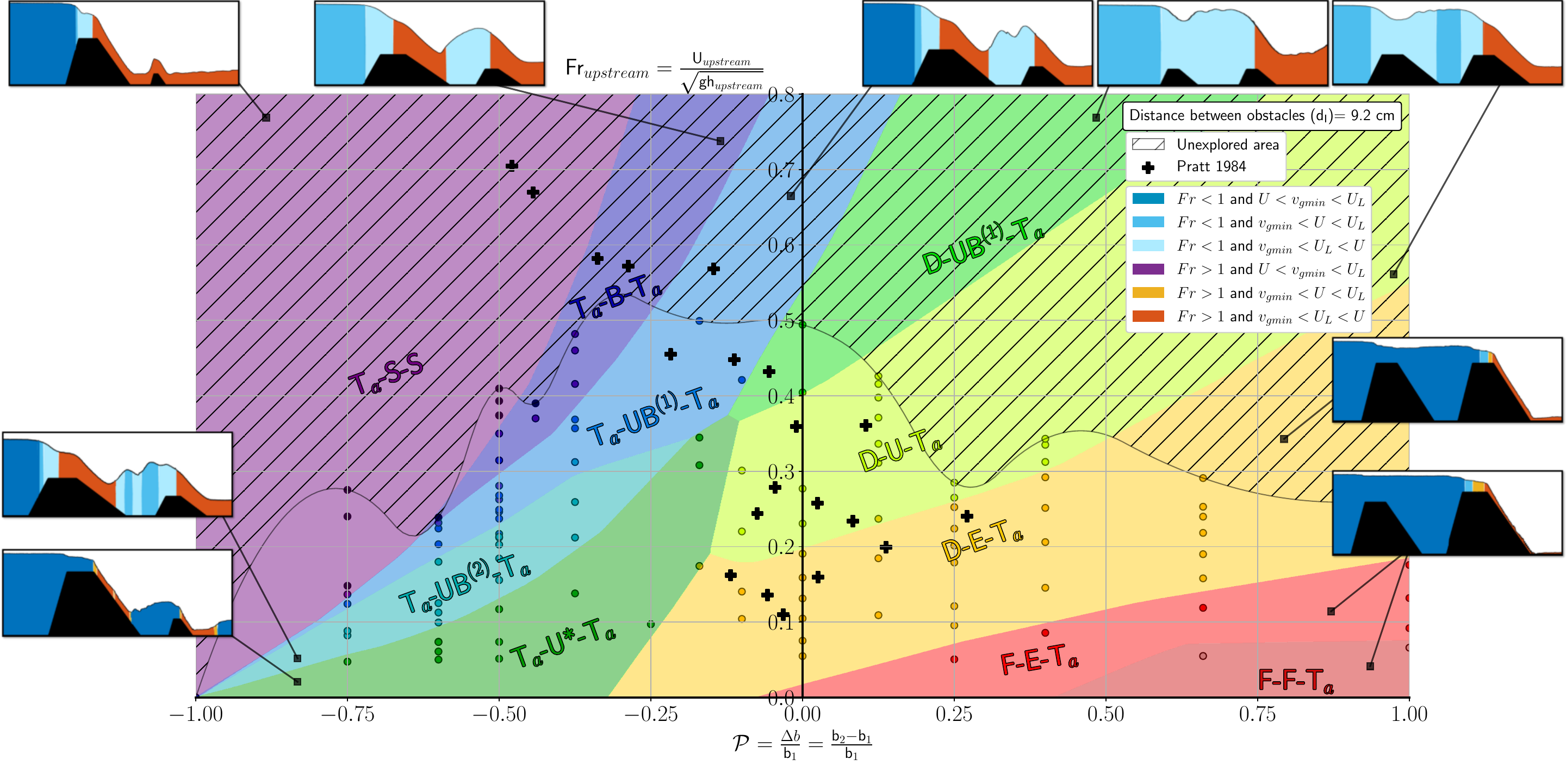}
\caption{Improved experimental Pratt's diagram, $Fr_\text{up}(\mathcal{P})$. The diagram is split into two parts by a hatched area that cannot be explored experimentally. The coloured circles are our experimental points. The black crosses are Pratt's 1984 experimental points. However, as these points are in a different plane, i.e. at a different distance between both obstacles, the Pratt points are only indicative. The points with coordinates (-1,0) are not experimental points but have been added because of the tendencies of the domains to converge towards this point. The figures around the edge of the diagram are not related to experimental points, they are there to illustrate what is expected in the field. The figures were obtained by an interface extraction technique (as in \cite{Weinfurtner-et-al-2011, Euve-et-al-2015, euve2016observation, Euve-Rousseaux-2017, euve2020scattering, fourdrinoy2022correlations}) on the experiments conducted. The different colour ranges in the images surrounding the diagram compare the modelled speed ($U=q/h$) with the minimum of the group speed, $v_\text{gmin}$, and the minimum of the phase speed, $U_L$ (that corresponds to the threshold controlled in hydrodynamics by the dispersive capillary length akin to a Planck scale where the Unruh analogy breaks down \cite{rousseaux2010horizon, rousseaux2020classical}) for the appearance of negative energy modes which is essential to both Hawking radiation and the black hole laser instability (see the Supplemental Material for details).
\label{fig:diagramm}}
\end{figure*}
\end{turnpage}
In the case of flows over two obstacles, several regimes are observed. To describe these regimes, a nomenclature is created based on the one introduced in \cite{rousseaux2020classical}, through a classification with a visual identification of the free surface shape. This nomenclature uses a succession of three symbols which characterise, respectively, the type of flow above the first obstacle, the type of flow between the two obstacles, and the type of flow above the second obstacle. The letters used for this nomenclature are: $S$ for Supercritical, $T_a$ for Transcritical (accelerating for black hole horizon), $B$ for Breaking, $U$ for Undular (the undulation is the zero frequency solution of the dispersion relation \cite{coutant2014undulations}), $D$ for Depression, $E$ for Emission and $F$ for Flat. Ten regimes are identified (without the use of an entrance sluice gate to accelerate the flow) and using the above nomenclature the names of the ten regimes are: $T_a$-$S$-$S$, $T_a$-$B$-$T_a$, $T_a$-$UB^{(2)}$-$T_a$, $T_a$-$UB^{(1)}$-$T_a$, $T_a$-$U^{*}$-$T_a$, $D$-$UB^{(1)}$-$T_a$, $D$-$U$-$T_a$, $D$-$E$-$T_a$, $F$-$E$-$T_a$ and $F$-$F$-$T_a$. Pictures of the regimes are available in the Supplemental Material (figure \ref{regimes}). The regimes with the ($i$) exponent are the regimes where the breaking regime of the undulation starts for the $(i+1)^\text{th}$ stubble (or whelp). For example, for the regime $T_a$-$UB^{(2)}$-$T_a$, the free surface between both obstacles is undulating ($U$) and the breaking regime ($B$) starts at the $3^\text{rd}$ stubble. The $U^*$-regime characterises the Undulation in shallow water and $U$ denotes the Undulation in deep water. Finally, the $E$-regime is a flat regime where there is a co-current and counter-current waves emission. Experimentally, these waves are emitted by a ``parabola'' which crosses the channel. The position of the parabola is interpreted as the place where the fluid velocity is approximately equal to the minimum of the group velocity \cite{rousseaux2010horizon, rousseaux2020classical}. 

The observed regimes are reported in a modernised Pratt diagram, Figure \ref{fig:diagramm}, which was constructed using deep learning tools (see the Supplemental Material). This diagram identifies new regimes that do not appear in the original diagram \cite{Pratt1984}, namely: $T_a$-$S$-$S$, $T_a$-$B$-$T_a$, $F$-$E$-$T_a$ and $F$-$F$-$T_a$. Pratt's experimental points are also superposed on it. However, it seems that the latter do not all correspond to the domains found. Indeed, Pratt uses another inter-obstacle distance not reported in his original work, which makes it corresponds to another plane in a three dimensional diagram, derived from a dimensionless number that would play the role of a third axis. We have not made this number explicit but we anticipate it to be a function of viscous friction, maybe a Reynolds number comparing the inter-obstacle distance and a viscous scale due to friction and head losses as it is known in the context of weir fishways \cite{Weirfish}. In the context of Analogue Gravity, the candidate regimes for the black hole LASER effect are: $T_a$-$B$-$T_a$, $T_a$-$UB^{(2)}$-$T_a$, $T_a$-$UB^{(1)}$-$T_a$, $T_a$-$U^{*}$-$T_a$ and $D$-$UB^{(1)}$-$T_a$ since these regimes have an undulation (a zero-frequency solution of the dispersion relation) with or without wave breaking and a speed allowing for the appearance of negative-energy waves \cite{rousseaux2010horizon, rousseaux2020classical}. Finally, when using a sluice gate in the upstream region, the acceleration of the upstream flow has made it possible to reach regimes that are not introduced in the original Pratt diagram \cite{Pratt1984}. These regimes are $S$-$S$-$S$ and $S$-$B$-$T_a$. A phase diagram that includes these two new regimes is shown in Figure~\ref{Diagramme de phase sluice gate} (see the Supplemental Material).


To summarize, we have studied here different flow regimes in a one-dimensional free surface water channel over one or two obstacles.  For a single obstacle, we experimentally validated Long's law for transcritical flows (without downstream boundary conditions and without initial static water height).  In Analogue Gravity, such flows correspond to an analogue ``black hole'', as there is a position where the local Froude number crosses 1.  We observed that Long's law tends to work better for large obstacles, because boundary layer effects are less important than for smaller obstacles; but even for the latter, the deviations from Long's law are relatively small. However, it appears that for the smallest obstacles and despite the viscous effects, the obstacles that best ``accompany the flow'' are those closest to the non-viscous Long law, such as the Orsay obstacle or the ACRI 2010 obstacle. For flows over two successive obstacles, we modernised and completed the Pratt diagram by observing several regimes, adding six to those that had already been described by Pratt. Among all regimes, five are potential candidates for the occurrence of the black hole LASER effect, a subject of future enquiry.

We would like to thank R. Bellanger, L. Dupuis, J.-C. Rousseau and R. Tessier for their help in the design of the experiments. We are indebted to E. Lamballais for the lending of the pedagogical water channel. The Master Thesis of A. Bossard was funded by the French government program ``Investissement d'Avenir'' (EUR INTREE, reference ANR-18-EURE-0010). This work pertains to the French government program ``Investissements d’Avenir'' (LABEX INTERACTIFS, reference ANR-11-LABX-0017-01). S. Robertson is funded through the CNRS Chair in Physical Hydrodynamics (reference ANR-22-CPJ2-0039-01).




\bibliography{biblio}


\section{Supplemental material}
\subsection{Experimental setup and metrology}

At the Pprime Institute, we have carried out experiments in an open water channel (reference H23 from Prodidac with transparent Plexiglas walls and an anodized aluminum support) to test the robustness of the Long's theory and its Faber approximation in the case of one obstacle. The channel dimensions (see figure \ref{exp-setup} top) are: length $L=2.5$ m, height $Z=12$ cm and width $W=5.3$ cm. It was also used to build the phase diagram, with two obstacles. The channel features a downstream guillotine to impose or not a backward flow. In all the following studies, no downstream condition is imposed, i.e. the downstream door remains wide open, there is no initial static water height and the flow is steady. An inclined honeycomb is placed upstream of the channel to laminarize locally the flow. The inclination of the honeycomb is approximately 66° to the bottom of the channel. The flow is sent through a pump upstream of the channel. This pump delivers a flow rate per unit width $q=Q/W$ with $q\in\mathclose{[}0.0006\,;0.0115\mathclose{]}$ $m^2.s^{-1}$. It is measured with a flowmeter Vortex F 20 (DN20) from Bamo Mesures. A Ultimaker 5S 3D printing machine is used with the CURA software with either polylactic acid (PLA) (black) or Acrylonitrile butadiene styrene (ABS) (blue) filaments of diameter 2.85 mm. Notches measuring 6 mm in width and 3.5 mm in depth allow to fix the obstacles on both sides of the channel. The free surface of the flow is illuminated by an overhead LED lighting system. The side meniscus is recorded by two grayscale (256) Point Grey cameras with CMOS technology. A MATLAB algorithm combines images from both cameras with a resultant recording length of 2.07 m with spatial resolution $dx = 0.5 mm$, during around 5 minutes at 25 fps for the acquisition rate. Then, the meniscus interface is processed with a subpixel detection method as in \cite{Weinfurtner-et-al-2011, Euve-et-al-2015, euve2016observation, Euve-Rousseaux-2017, euve2020scattering, fourdrinoy2022correlations}. We detect first the maximum intensity corresponding to the meniscus on the combined images for each position x and each time t. The aberrant points caused by blurs or water drops are substituted by an average value of the point neighborhood. A second step consists in seeking the maximum brightness around the positions previously found culminating with a one pixel precision. Finally, a sub-pixel detection method allows to fit with a Gaussian over five points around the maximum value in the vertical direction of the luminosity reaching a fraction of the pixel size $dx$. One purpose of detection is shown in the figure \ref{exemple extraction}. In this figure, the water level is shown as a solid black line. A black mask is applied to locate the obstacle. Coloured areas appear in the detection corresponding to whether the local Froude number is greater than 1 (red area) or less than 1 (blue area). Within these zones, there are coloured sub-zones that compare the fluid velocity, $U(x)=q/h(x)$, to the minimum of the group velocity, $c_\text{gmin}$, and the minimum of the phase velocity, $U_L$ (the threshold for the appearance of negative energy waves as discussed in \cite{rousseaux2010horizon}). $c_\text{gmin}$ and the so-called Landau speed $U_L$ are calculated from the following dispersion relation (see \cite{rousseaux2010horizon, rousseaux2020classical} for details):

\begin{linenomath}
\begin{equation}
\omega^2=gk\left(1+\frac{\gamma}{\rho g}k^2\right)\tanh\left(kh\right)
    \label{eq:disp}
\end{equation}
\end{linenomath}

\begin{appendices}
\begin{figure*}
\includegraphics[width=0.95\textwidth]{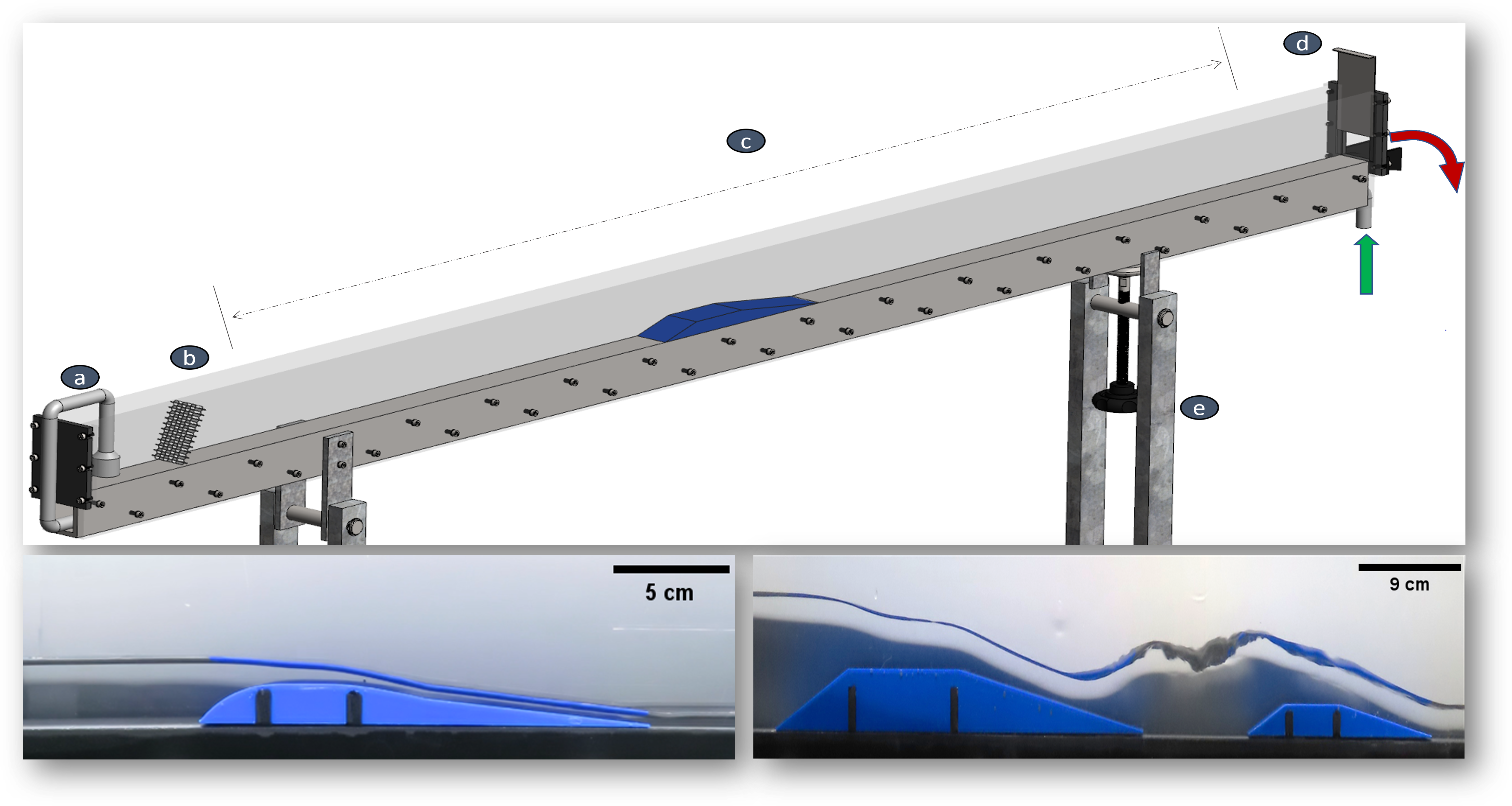}
\caption{Top: This image is a CAD (Computer-aided design) of the channel we used for our experiments. The flow is sent by the pump into the channel (green arrow) and arrives at its entrance from above through a nozzle at point a).  The turbulent flow passes through a honeycomb b), to laminarize it locally. The test section, c), is about 2.1m long. It is in this area that the detection and observations are made. Finally, the flow passes through a guillotine, d), and is re-injected into the pump (red arrow) through a water fall with a disconnected exit chamber. Thanks to the knob e), the channel is placed at the horizontal.
Bottom left: illustrative photo of a flow over one obstacle. The flow goes from left to right over a blue ABS obstacle. The obstacle is Vancouver 0.11.
Bottom right: illustrative photo of a flow over two successive obstacles. The flow goes from left to right over the blue ABS obstacles. The obstacles are ACRI 2010 x0.02 and ACRI 2010 x0.01 inspired from the one used in \cite{chaline2013some}.
\label{exp-setup}}
\end{figure*}

\subsection{Influence of obstacles sizes and geometries}
\begin{figure*}
\includegraphics[width=0.95\textwidth]{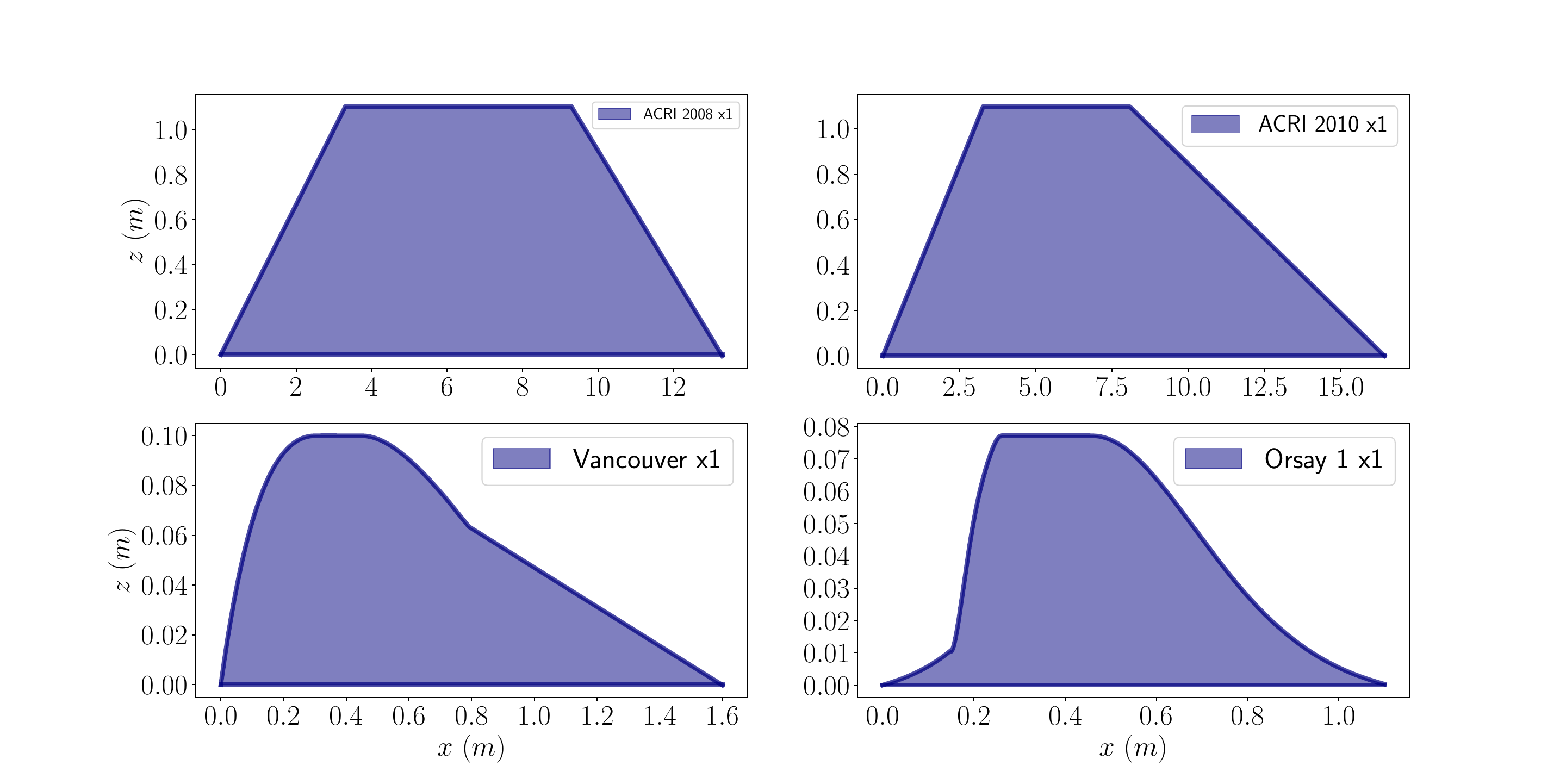}
\caption{Profiles of the obstacles used. Homothety ratios were applied from these obstacles. Image in the top left corner: ACRI 2008 x1 obstacle \cite{Rousseaux-et-al-2008, rousseaux2010horizon}; Image in the top right corner: ACRI 2010 x1 obstacle \cite{chaline2013some}; Image in the bottom left corner: Vancouver x1 \cite{Weinfurtner-et-al-2011}; Image in the bottom right corner: Orsay 1 x1, an obstacle designed by Florent Michel and Renaud Parentani in the same vein as the Orsay 2 obstacle we used in \cite{euve2016observation}.
\label{exp-obstacles}}
\end{figure*}

To test the robustness of the theories around the flow over an obstacle, we used a panel of obstacles of different sizes and geometries, as can be seen in figure \ref{exp-setup} and figure \ref{exp-obstacles}, where the bottom left image is an obstacle with a more "aerodynamic" geometry (which we call Vancouver geometry as in \cite{Weinfurtner-et-al-2011}), or on the bottom right image, where we use a different geometry and different sizes; the obstacles of this image is called ACRI 2010 as in \cite{chaline2013some}.
All obstacles are built with a 3D printer in ABS (acrylonitrile butadiene styrene). The obstacles are built from a reference model where a homothety is applied on all the lengths of the obstacles except the width W, because the very same width is needed to insert all the obstacles in the water channel.

\begin{figure*}
\includegraphics[width=0.95\textwidth]{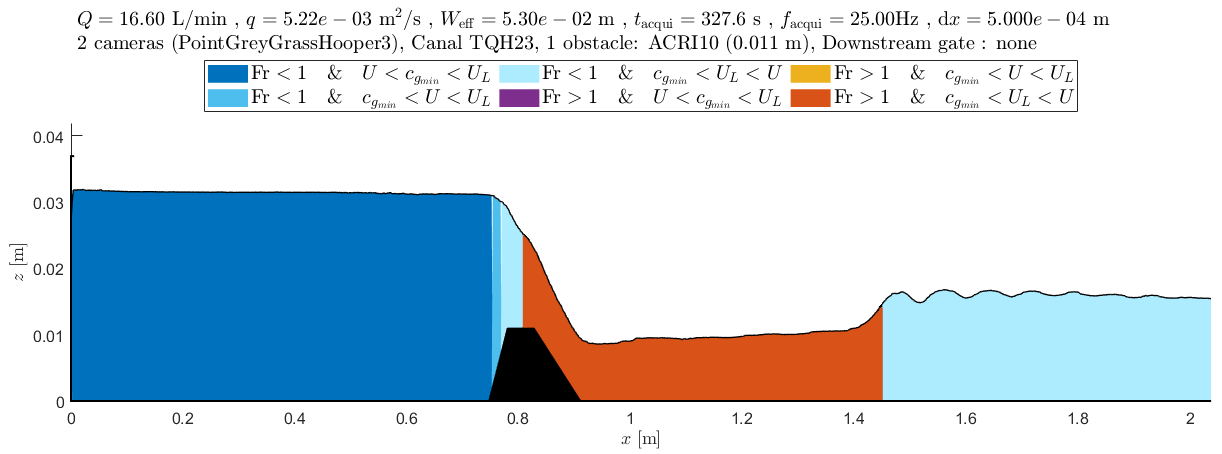}
\caption{Example of interface extraction for an obstacle, here the ACRI 2010 x0.01 as in \cite{chaline2013some}, without downstream condition and static water depth.
\label{exemple extraction}}
\end{figure*}

\begin{figure*}
\includegraphics[width=0.95\textwidth]{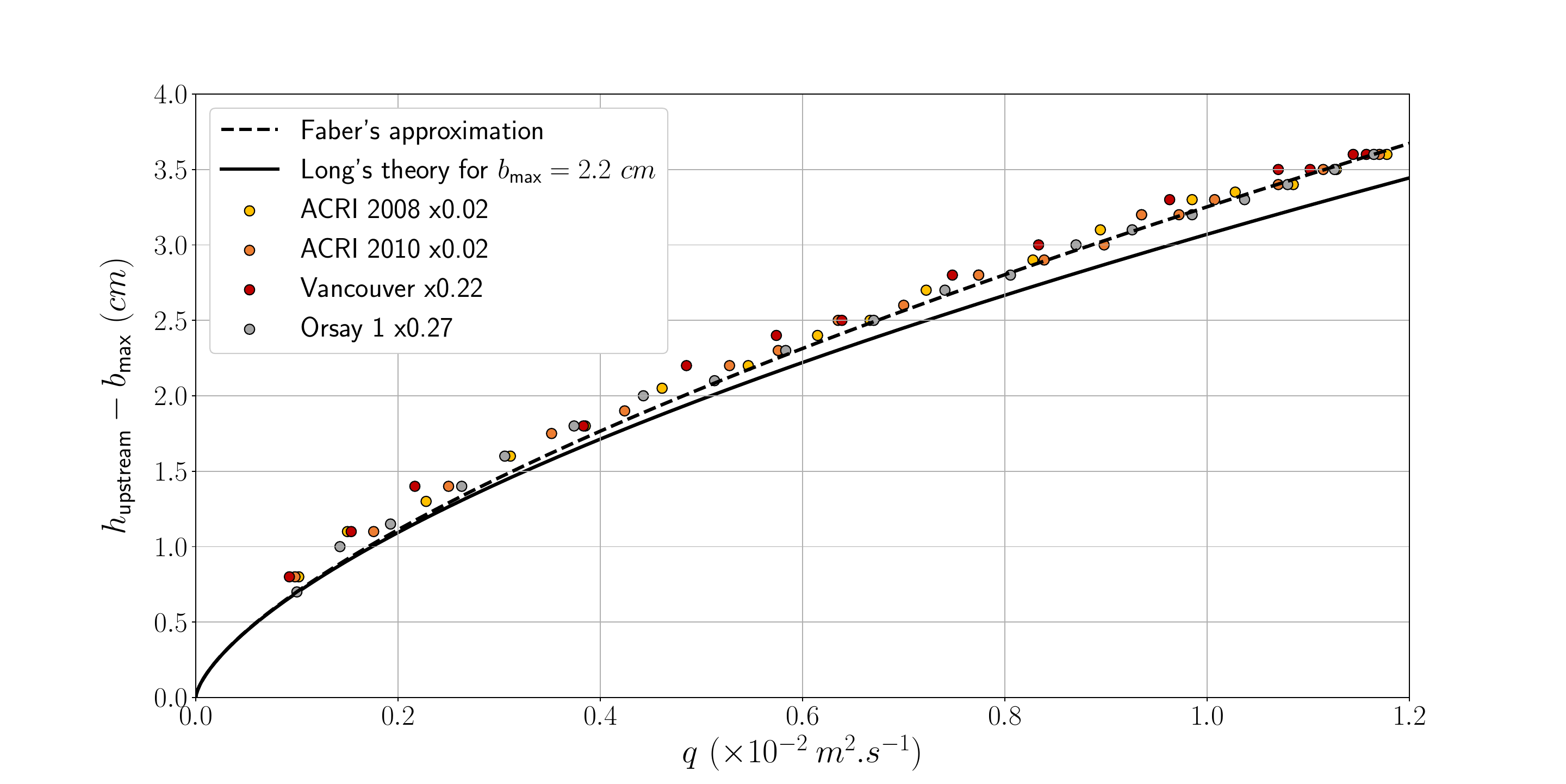}
\caption{Influence of the obstacle geometry compared to Long's theory \cite{Long}. This graph was obtained in the channel of figure \ref{exp-setup}, without downstream condition and without static water depth.
\label{comparaison camille shape}}
\end{figure*}

From an experimentalist's point of view, the $Fr_\text{up}(r)$ diagram is replaced by the diagram $\left(h_\text{upstream}-b_\text{max}\right)(q)$.  It is in this diagram that the studies on the geometry and size of the obstacles were done separately in practice. By changing the geometry and setting the maximum size of the obstacles, we obtain the figure \ref{comparaison camille shape}. The experimental points show that these obstacle geometries tend to follow the Faber approximation (black hatched line) and are in a transcritical regime; even though the points are above the theoretical Long curve (black solid line), which means that, theoretically, they are subcritical regimes.

\begin{figure*}
\includegraphics[width=0.95\textwidth]{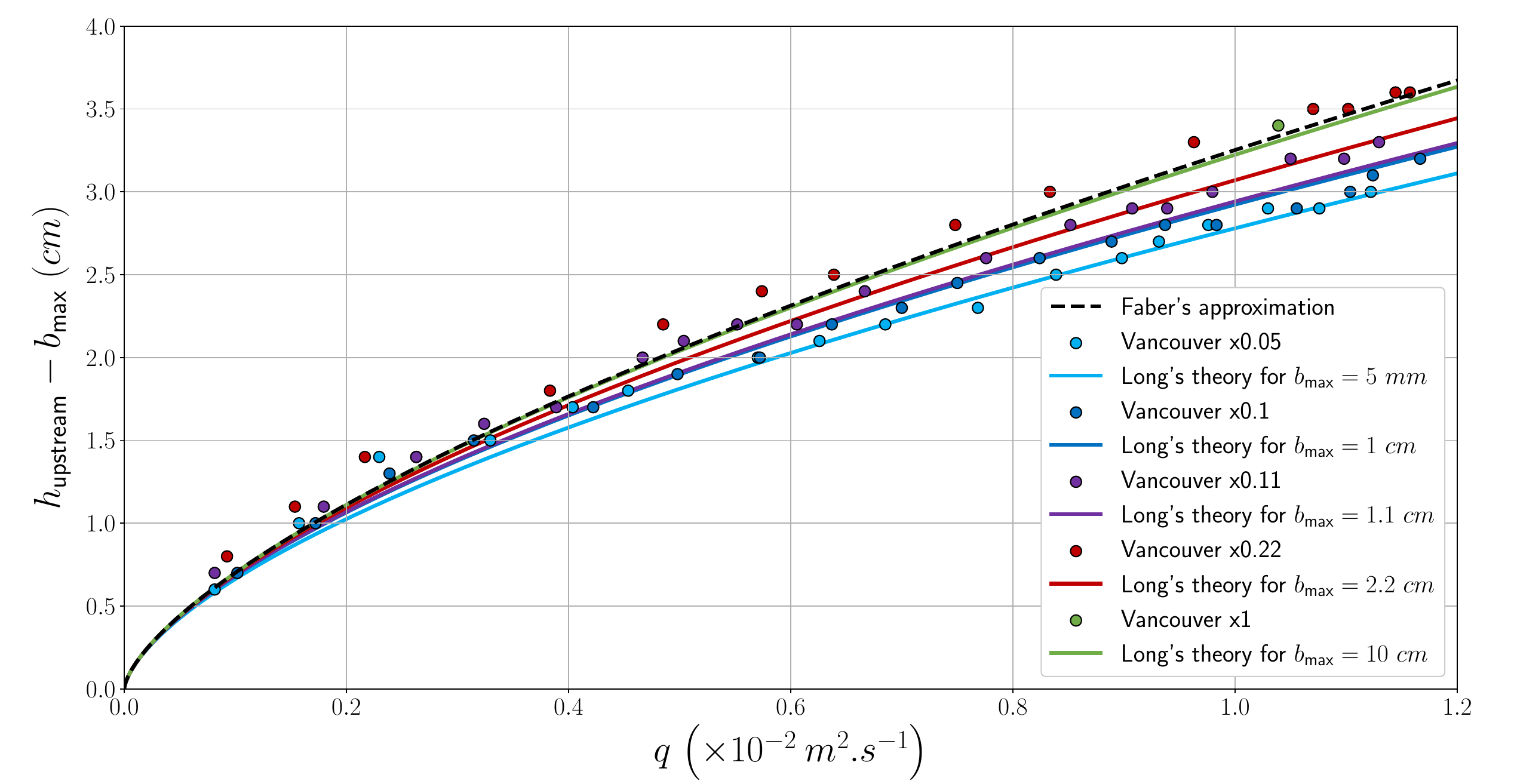}
\caption{Influence of the obstacle size for a given geometry. In this case, the geometry used is the one of Vancouver \cite{Weinfurtner-et-al-2011}, an example of which can be seen on the bottom left of figure \ref{exp-setup}. This graph was obtained in the channel of figure \ref{exp-setup}, without downstream condition and without static water depth.
\label{comparaison camille scale}}
\end{figure*}

On the other hand, for the figure \ref{comparaison camille scale} the geometry has been set to the Vancouver type and the size of the obstacles is variable. According to the figure \ref{comparaison camille scale}, small obstacles, such as the Vancouver x0.05, converge towards Long's law when the flow rate increases. Otherwise, the obstacle Vancouver x0.22 is furthest from the long theory for a certain height of 2.2 cm, when the flow rate increases.

\begin{figure*}
\includegraphics[width=0.95\textwidth]{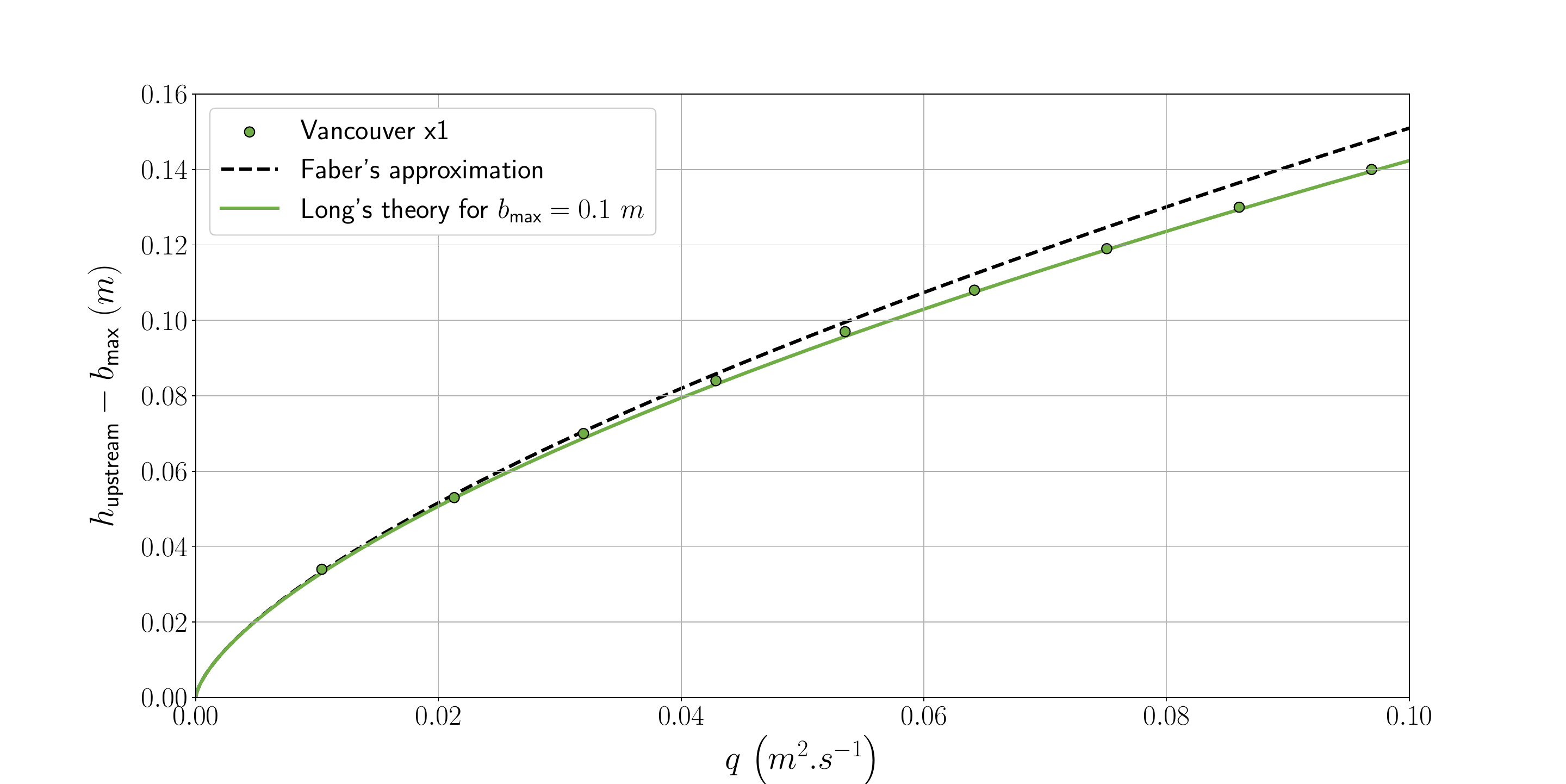}
\caption{This graph was obtained by taking a larger Vancouver-type obstacle with a length of 1.6 m and a maximum height of 10 cm placed in a channel 6 m long and 38.5 cm wide as in \cite{Euve-et-al-2015, euve2016observation, Euve-Rousseaux-2017, euve2020scattering}. This graph was obtained without downstream conditions and without static water depth.
\label{camille vancouver 1}}
\end{figure*}

To display the Long's theory in the graph $h_\text{upstream}-b_\text{max}$ as a function of q (solids lines on the graphics \ref{comparaison camille shape}, \ref{comparaison camille scale} and \ref{camille vancouver 1}), one must take the Long's law (equation \ref{Long}) and isolate the variable $Y=h_\text{upstream}-b_\text{max}$. This gives the following third degree polynomial verified by Y:
\begin{linenomath}
\begin{multline}
    Y^3+\left(2b_\text{max}-\frac{3}{2}\sqrt[3]{\frac{q^2}{g}}\right)Y^2+\left(b_\text{max}^2-3\sqrt[3]{\frac{q^2}{g}}b_\text{max}\right)Y \\
    +\frac{1}{2}\frac{q^2}{g}-\frac{3}{2}\sqrt[3]{\frac{q^2}{g}}b_\text{max}^2=0
    \label{eq:long pour graphique camille}
\end{multline}
\end{linenomath}
where $b_\text{max}$ is the maximum height of the obstacle and $g=9.81$ $m.s^{-2}$. By solving the polynomial, we find that :

\begin{linenomath}
\begin{multline}
h_\text{upstream}=\frac{1}{2}\sqrt[3]{\frac{q^2}{g}}+\frac{b_\text{max}}{3}+\\
2\left(\frac{1}{2}\sqrt[3]{\frac{q^2}{g}}+\frac{b_\text{max}}{3}\right)\times\\
\cos\left(\frac{1}{3}\Arccos\left(1-\frac{\frac{1}{4}\frac{q^2}{g}}{\left(\frac{1}{2}\sqrt[3]{\frac{q^2}{g}}+\frac{b_\text{max}}{3}\right)^3}\right)\right)
    \label{hamont}
\end{multline}
\end{linenomath}

The Faber approximation \ref{Faber} can be derived from this long-dimensional law (equation \ref{hamont}) by developing it asymptotically as the ratio $h_c/b_\text{max}$ is very small compared to 1 where $h_c=\sqrt[3]{q^2/g}$ is the critical water height, i.e. the height of the fluid when the Froude number is equal to 1 \cite{Long} (it corresponds to the first transition from cyan to red colours in the figure \ref{exemple extraction}).

\begin{figure*}
\includegraphics[width=0.95\textwidth]{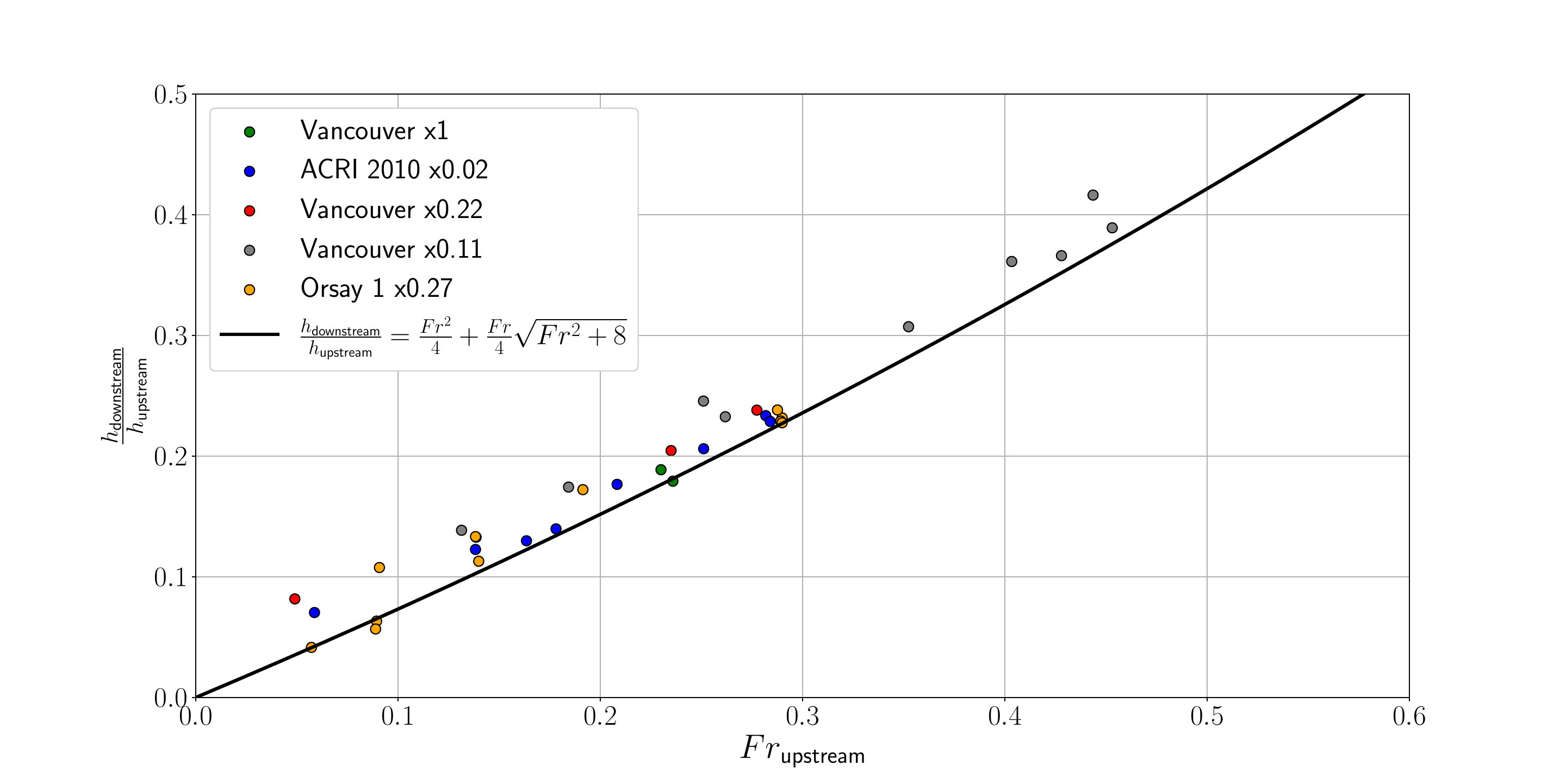}
\caption{Experimental and theoretical graph of the ratio of the downstream water depth by the upstream water depth as a function of the Froude number. This graph was obtained without downstream conditions and without static water depth.
\label{h_d sur h_u}}
\end{figure*}

\begin{figure*}
\includegraphics[width=0.95\textwidth]{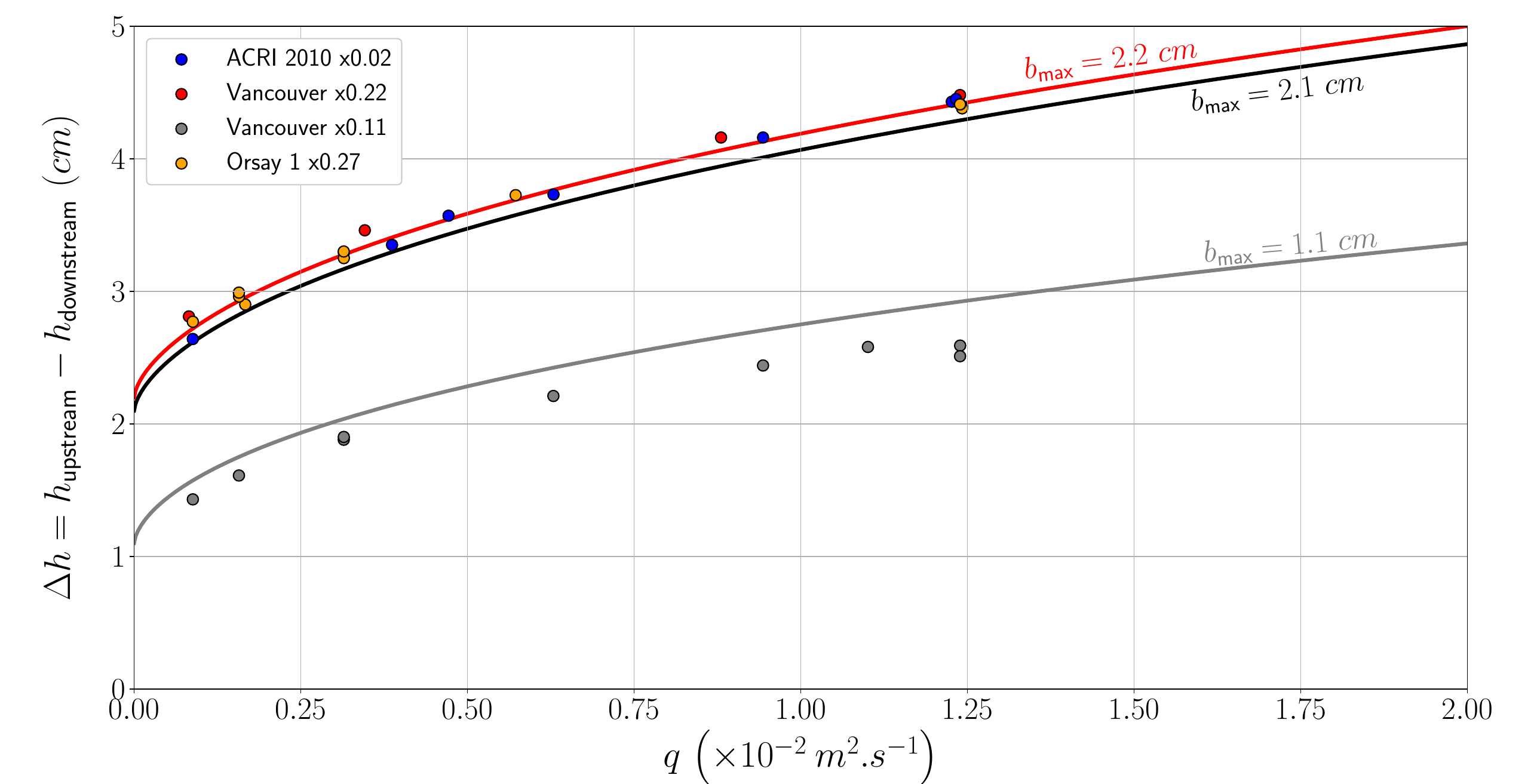}
\caption{Evolution of the height of the transcritical flow cascade as a function of the flow rate. The solid lines are from Long's theory \cite{Long} and in this case if $q=0$ then $\Delta h=b_\text{max}$.This graph was obtained without downstream conditions and without static water depth. Furthermore, by asymptotic analysis, we show that: $h_\text{upstream}-h_\text{dowstream}\underset{q\rightarrow +\infty}{\sim}\frac{2\sqrt{6}}{3}\frac{\sqrt{b_\text{max}}}{\sqrt[6]{g}}\sqrt[3]{q}$. Thus, according to Long's model, the height of the transcritical cascade tends to $+\infty$ as q tends to $+\infty$. Indeed, the asymptotic development of $h_\text{upstream}$ is : $h_\text{upstream}=\frac{1}{\sqrt[3]{g}}q^{\frac{2}{3}}+\frac{\sqrt{6}\sqrt{b_\text{max}}}{3\sqrt[6]{g}}q^{\frac{1}{3}}+\underset{q\rightarrow +\infty}{O}\left(1\right)$
\label{deltah(q)}}
\end{figure*}

\subsection{Deviations from Long's formula due to boundary friction}

Long's formula for a transcritical flow can be derived from the stationary Saint-Venant equations:
\begin{eqnarray}
\partial_{x}\left[ u(x) h(x) \right] &=& 0 \,, \nonumber \\
\partial_{x}\left[ \frac{u^{2}(x)}{2} + g (h(x) + b(x)) \right] &=& 0 \,.
\label{eq:saint-venant}
\end{eqnarray}
The first of these equations encodes mass conservation, and allows us to write $u(x) = q/h(x)$ where $q$ is the flow rate per unit width.  The second equation represents energy conservation.

To take account of the effect of boundary friction, we leave the first of Eqs.~(\ref{eq:saint-venant}) as it is, and modify the second to allow for a small dissipation of the energy:
\begin{equation}
\partial_{x}\left[ \frac{u^{2}(x)}{2} + g (h(x) + b(x)) \right] = -f(u) \,,
\label{eq:saint-venant_with_friction}
\end{equation}
where $f(u)$ is some positive function of the velocity $u$ that gives the rate of energy loss per unit distance {\it per unit mass} in the direction of the flow.  Here we shall assume a straightforward linear function of the velocity: $f(u) = \sigma u$ for some constant $\sigma$ that is a property of the particular flume being used.  Note that, as the energy loss rate is given per unit mass, we can expect $\sigma$ to be larger for narrower flumes.  The occurrence of back-flow due to the presence of the obstacle also seems to play an important role in the determination of $\sigma$.  From an examination of the flows achieved in the absence of an obstacle, we estimate that, for the narrow flume used in the experiments, we have $\sigma \approx 0.07 \, s^{-1}$; whereas, in the upstream region in the presence of an obstacle, the effective $\sigma$ is an order of magnitude smaller than this.

In any case, for a given $\sigma$ and obstacle profile $b(x)$, we get an explicit expression for the rate of change of the water depth $h(x)$:
\begin{equation}
    h^{\prime}(x) = \frac{-g b^{\prime}(x) h^{3}(x) - \sigma q h^{2}(x)}{g h^{3}(x) - q^{2}} \,.
    \label{eq:hprime}
\end{equation}
At the horizon, $u^2 = q^2/h^2 = g h$, which occurs exactly where the denominator on the RHS of~(\ref{eq:hprime}) vanishes.  To avoid any divergence, the numerator must simultaneously vanish, which means that at the horizon we must have
\begin{equation}
    b^{\prime} = -\frac{\sigma q}{g h} = -\sigma \sqrt{\frac{h}{g}} \,.
\end{equation}
Unlike in Long's frictionless scenario, where the horizon must occur where $b^{\prime} = 0$, here we find that the horizon is pushed towards the downstream side of the obstacle where $b$ is decreasing.  We also find that the position of the horizon (which determines $b^{\prime}$ for a given obstacle), the water depth $h$ at the horizon, and the flow rate $q$ are not independent quantities: fixing one determines the other two.

Similarly, $h^{\prime}$ is also determined at the horizon, for the vanishing of numerator and denominator at the same point means that we can apply l'H\^{o}pital's rule to Eq.~(\ref{eq:hprime}).  This yields a quadratic equation for $h^{\prime}$:
\begin{equation}
    3 g h^{2} (h^{\prime})^{2} + (3 g h^{2} b^{\prime} + \sigma q^{2}) h^{\prime} + g h^{3} b^{\prime\prime} = 0 \,.
\end{equation}
In fact there are two possible values of $h^{\prime}$ for each $q$.  These correspond to an accelerating and a decelerating flow at the horizon, i.e., to a black-hole and a white-hole flow.

\begin{figure}
    \centering
    \includegraphics[width=0.95\columnwidth]{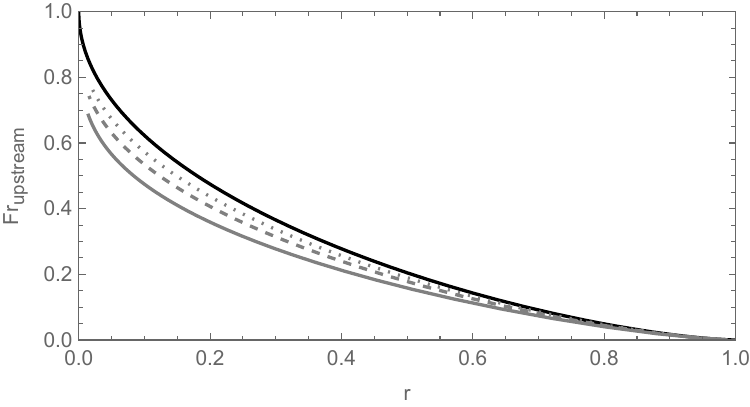}
    \caption{Deviations from Long's formula \cite{Long}.  The black solid line is the Long curve relating the obstruction ratio $r$ to the upstream Froude number in the absence of friction.  Using Eqs.~(\ref{eq:saint-venant_with_friction})-(\ref{eq:initialization_of_friction_case}), we find deviations like the three gray curves, which are found here for fixed $\sigma$ but at various distances $d$ upstream from the obstacle.  We find almost exactly the same deviations if the distance is fixed and $\sigma$ is varied, such that the product $\sigma d$ seems to be the relevant parameter that governs the deviations.}
    \label{fig:Long_dissipation}
\end{figure}

We can solve the equations numerically for different $q$ by determining the associated position of the horizon, as well as $h$ and $h^{\prime}$ at the horizon.  We may then give, as initial conditions a short distance $\epsilon$ away from the horizon,
\begin{equation}
    h(x+\epsilon) = h_{0} + h^{\prime}_{0} \, \epsilon \,,
    \label{eq:initialization_of_friction_case}
\end{equation}
and proceed to integrate Eq.~(\ref{eq:hprime}) out into the asymptotic region.  In order to extract an upstream water depth (for the obstruction ratio $r = b_{max}/h_{upstream}$) and the upstream Froude number $Fr_{up} = u_{upstream}/c_{upstream}$, we need to choose a location at which these quantities are evaluated, since the friction causes them to vary with position.  So we get a family of curves in the $(r,Fr_{up})$ plane, examples of which are shown in Fig.~\ref{fig:Long_dissipation}.

The results of this section show that friction can account for the deviations from Long's formula in a quantitative sense: it tends to push the data points downwards in the $(r,Fr_{up})$ plane, just as we observe for most of the data points shown in Fig.~\ref{fig:camille}.  As noted in the caption of Fig.~\ref{fig:Long_dissipation}, we find that it is the product $\sigma d$ that determines the form of the deviations, and that a value of $\sigma d \approx 0.07 \, m/s$ tends to fit the experimental data points of Fig.~\ref{fig:camille} fairly well, particularly at lower values of $Fr_{up}$.  However, as the measurements of the upstream Froude number and water depth were performed fairly close to the obstacle at $d = 5 \, {\rm cm}$, this would seem to require a rather large friction coefficient $\sigma \approx 1.4 \, {\rm s}^{-1}$.  A more sophisticated analysis, including the effect of back-flow due to the obstacle and taking account of the vertical dependence of the fluid flow, may provide more quantitative understanding of these phenomena.

\subsection{Construction of the phase diagram}
\begin{figure}[h!]
\includegraphics[width=0.45\textwidth]{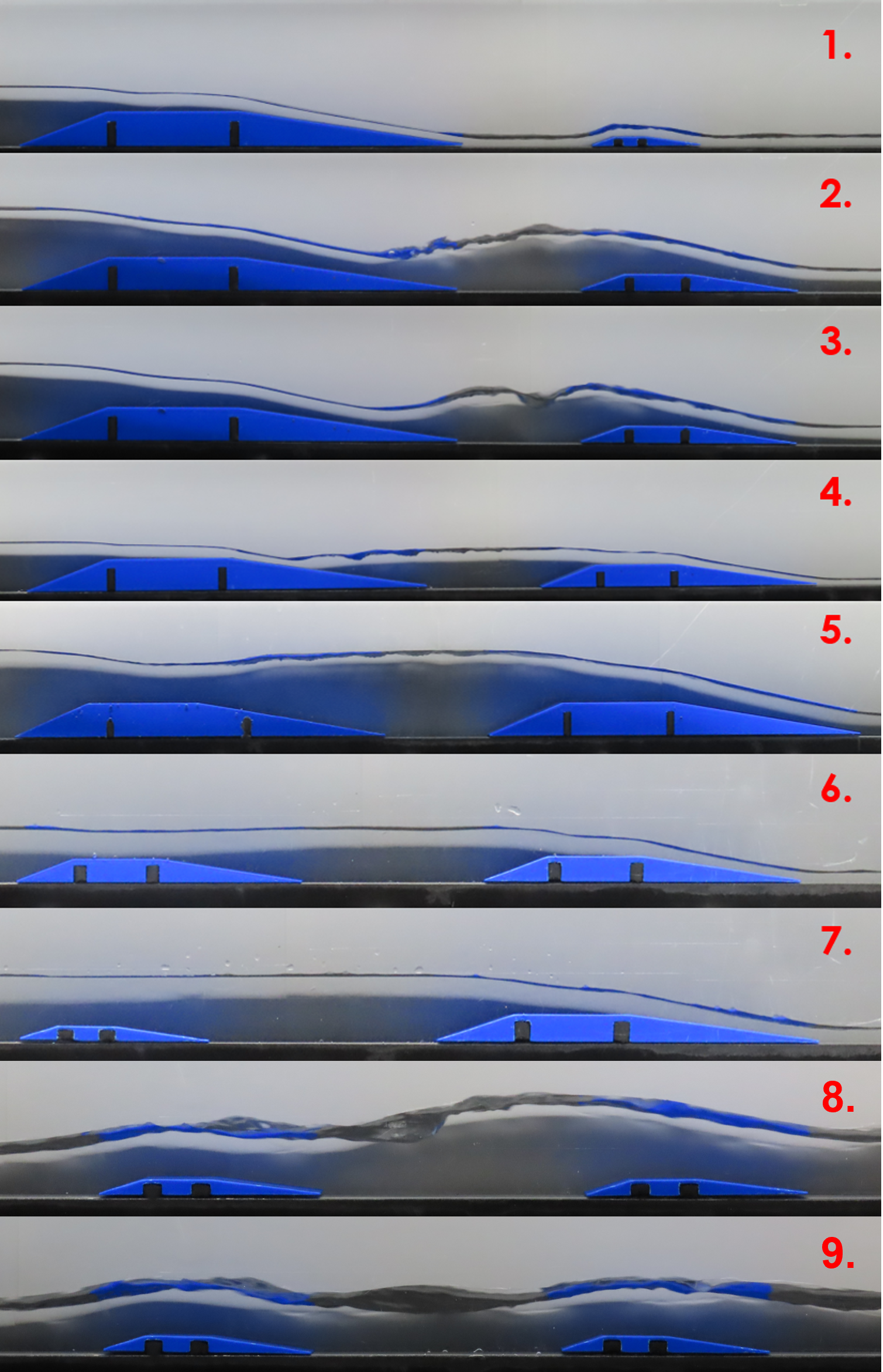}
\caption{Photos of the different regimes identified: 1. $T_a$-$S$-$S$; 2. $T_a$-$B$-$T_a$; 3. $T_a$-$UB^{(1)/(2)}$-$T_a$; 4. $T_a$-$U^{*}$-$T_a$; 5. $D$-$U/UB^{(1)}$-$T_a$; 6. $D$-$E$-$T_a$; 7. $F$-$F/E$-$T_a$; 8. $S$-$B$-$T_a$ and 9. $S$-$S$-$S$
\label{regimes}}
\end{figure}

For the construction of the diagram, the obstacles were placed in the channel with a separation distance of 9.2 cm. This distance was chosen arbitrarily as a result of various tests. This distance is measured from the end of the first obstacle to the beginning of the second. To construct a point in the phase diagram, the pump flow rate is set to the desired flow rate, always increasing the flow rate (to avoid hysteresis phenomena). Thus, the Pratt number is given by the maximum height of the two obstacles and the global upstream Froude number is formed with the measurement of the water depth, upstream of the first obstacle, $h_\text{upstream}$, and the pump flow rate, $Q$ (assuming the velocity profile $U_\text{upstream}=q/h_\text{upstream}$ with $q=Q/W$). Picture of the regimes are shown in the figure \ref{regimes}.

\subsubsection{The neural network}
\begin{figure}[h]
\def\layersep{2cm}
\centering
\begin{tikzpicture}[shorten >=1pt,->,draw=black!50, node distance=\layersep]
    \tikzstyle{every pin edge}=[<-,shorten <=1pt]
    \tikzstyle{neuron}=[circle,fill=white!25,minimum size=17pt,inner sep=0pt]
    \tikzstyle{input neuron}=[neuron, fill=green!50];
    \tikzstyle{output neuron}=[neuron, fill=red!50];
    \tikzstyle{hidden neuron}=[neuron, fill=blue!50];
    \tikzstyle{annot} = [text width=4em, text centered]

    
    \node[input neuron, pin=left:Input \#1] (I-1) at (0,-1 cm) {};
    \node[input neuron, pin=left:Input \#2] (I-2) at (0,-2 cm) {};  
    \node[input neuron, pin=left:Input \#n-1] (I-3) at (0,-3 cm) {};
    \node[input neuron, pin=left:Input \#n] (I-4) at (0,-4 cm) {};
    
    \fill[black] (0,-2.35) circle[radius=1pt];
    \fill[black] (0,-2.5) circle[radius=1pt];
    \fill[black] (0,-2.65) circle[radius=1pt];
    
    \fill[black] (-1,-2.35) circle[radius=1pt];
    \fill[black] (-1,-2.5) circle[radius=1pt];
    \fill[black] (-1,-2.65) circle[radius=1pt];

    \foreach \name / \y in {1,...,5}
        \path[yshift=0.5cm]
            node[hidden neuron] (H-\name) at (\layersep,-\y cm) {};

    \foreach \name / \y in {1,...,4}
        \path[yshift=0.01cm]
            node[output neuron,pin={[pin edge={->}]right:Output}, right of=H-3] (G-\name) at (\layersep,-\y cm) {};


    \foreach \source in {1,...,4}
        \foreach \dest in {1,...,5}
            \path (I-\source) edge (H-\dest);

	\foreach \source in {1,...,5}
        \foreach \dest in {1,...,4}
            \path (H-\source) edge (G-\dest);

    \node[annot,above of=H-1, node distance=1.5cm] (hl) {Hidden layer with 100 neurons};
    \node[annot,left of=hl] {Input layer with n neurons};
    \node[annot,right of=hl] {Output layer with k neurons};
    
    \fill[white] (2,-2.5) circle[radius=10pt];
    \fill[black] (2,-2.2) circle[radius=1pt];
    \fill[black] (2,-2.5) circle[radius=1pt];
    \fill[black] (2,-2.8) circle[radius=1pt];
    
    \fill[black] (5,-2.35) circle[radius=1pt];
    \fill[black] (5,-2.5) circle[radius=1pt];
    \fill[black] (5,-2.65) circle[radius=1pt];
\end{tikzpicture}
\caption{Architecture of the neural network with n the number of points in the training set and k the number of groups (here the number of hydrodynamic regimes).}
\label{deeplearning}
\end{figure}
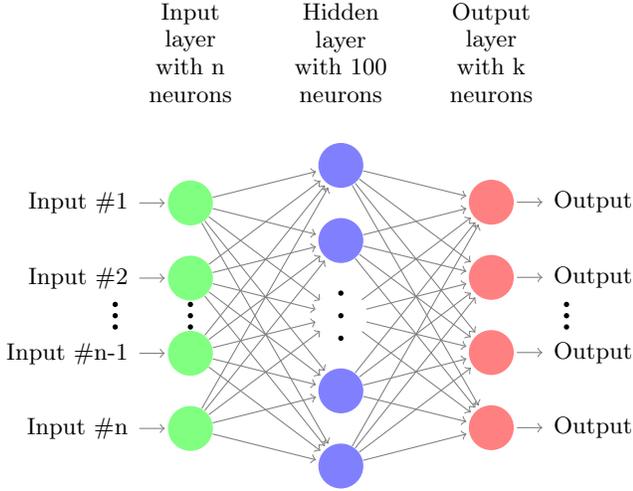

The experimental phase diagram, figure \ref{fig:diagramm}, was a classification problem that was solved using a deep learning approach. A neural network, whose schematic is given in figure \ref{deeplearning}, created with a python program, using the ScikitLearn library and the MLPClassifier function. It is a simple network, composed of a single hidden layer, itself composed of 100 neurons. The number of neurons in the input layer corresponds to the experimental points (the training set). All experimental points are assigned a label (a number corresponding to the regime to which they belong). The assignment of a regime to a point was done by visual identification of the free surface. The output layer of the network is composed of as many neurons as there are hydrodynamic regimes.

The loss function used by the MLPClassifier function is:
\begin{linenomath}
\begin{equation}
    L_\text{loss}=-\frac{1}{n}\displaystyle\sum_{i=1}^{n}\sum_{j=1}^{k}y_{i,j}\ln(p_{i,j}) + \frac{\alpha}{2n}\left\|W\right\|^2_2
\end{equation}
\end{linenomath}
where n the number of points and k the label of the regimes.
In this case, $y_{i,j}=1$ if point i has label j, and $p_{i,j}=\mathbb{P}\left(y_{i,j}=1\right)$ with the following property: $ \sum_{j=1}^kp_{i,j}=1$, for all i. The goal is to minimize the loss function and thus to find for a given point (i.e. for a given i) which j maximizes $p_{i,j}$. We can add a regularization term L2, but it is very weak, because it is multiplied by $\alpha=0.0001$, by default.
For example, for the figure \ref{fig:diagramm}, $n=119$ and $k=10$.

\subsubsection{Improvement of the diagram with a sluice gate}
\begin{figure*}
\includegraphics[scale=0.52,width=1.05\textwidth]{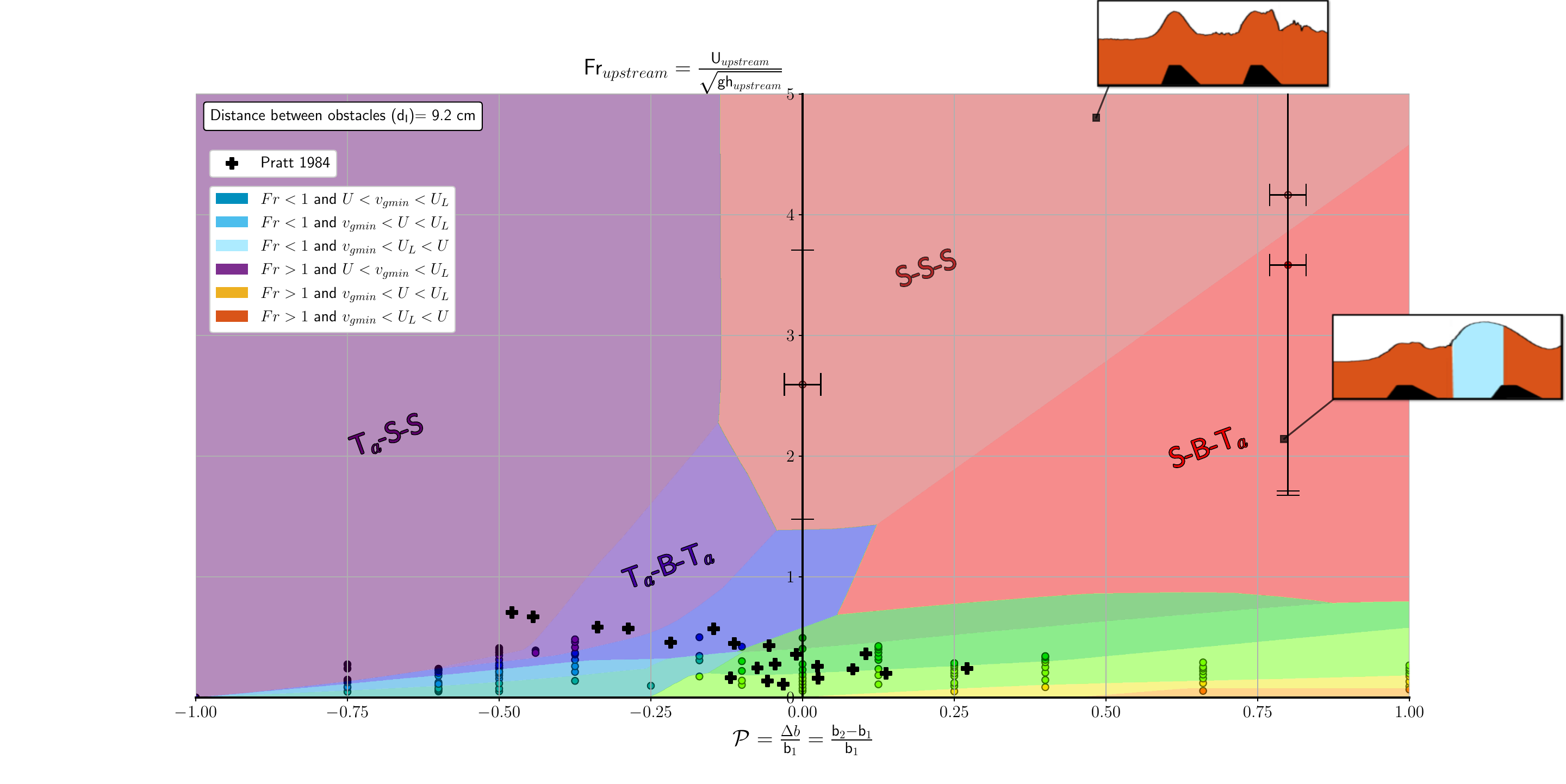}
\caption{Phase diagram with two obstacles, obtained without downstream condition and without initial static water height. This phase diagram completes the diagram \ref{fig:diagramm} with new regimes obtained with a lock gate positioned upstream of the first obstacle. The new regimes are $S$-$S$-$S$ and $S$-$B$-$T_a$. The uncertainties associated with the experimental points are larger, because the water depth and velocity flow upstream of the first obstacle are not constant, due to the pressure losses. These regimes have been identified with the obstacles ACR1 2010 x0.005 and ACRI 2010 x0.005 (in black on the interface extraction images, images that surround the diagram)
\label{Diagramme de phase sluice gate}}
\end{figure*}

In order to explore the phase diagram, figure \ref{Diagramme de phase sluice gate}, for higher Froude number values, a sluice gate is used, and placed before the honeycomb, to accelerate the flow. This sluice gate is inserted vertically (and remains vertical for all experiments). The distance between the channel bottom and the sluice gate is between $\mathclose{[}1.3;1.8\mathclose{]}$ cm. These new experiments with the sluice gate allow to extend the phase diagram and to propose 2 new regimes: S-B-T and S-S-S (figure \ref{Diagramme de phase sluice gate}).The uncertainties for these points are higher, because the Froude number is not spatially constant (due to pressure drops) upstream of the first obstacle.

\subsubsection{Uncertainties}
\begin{figure*}
\includegraphics[scale=0.52,width=1.05\textwidth]{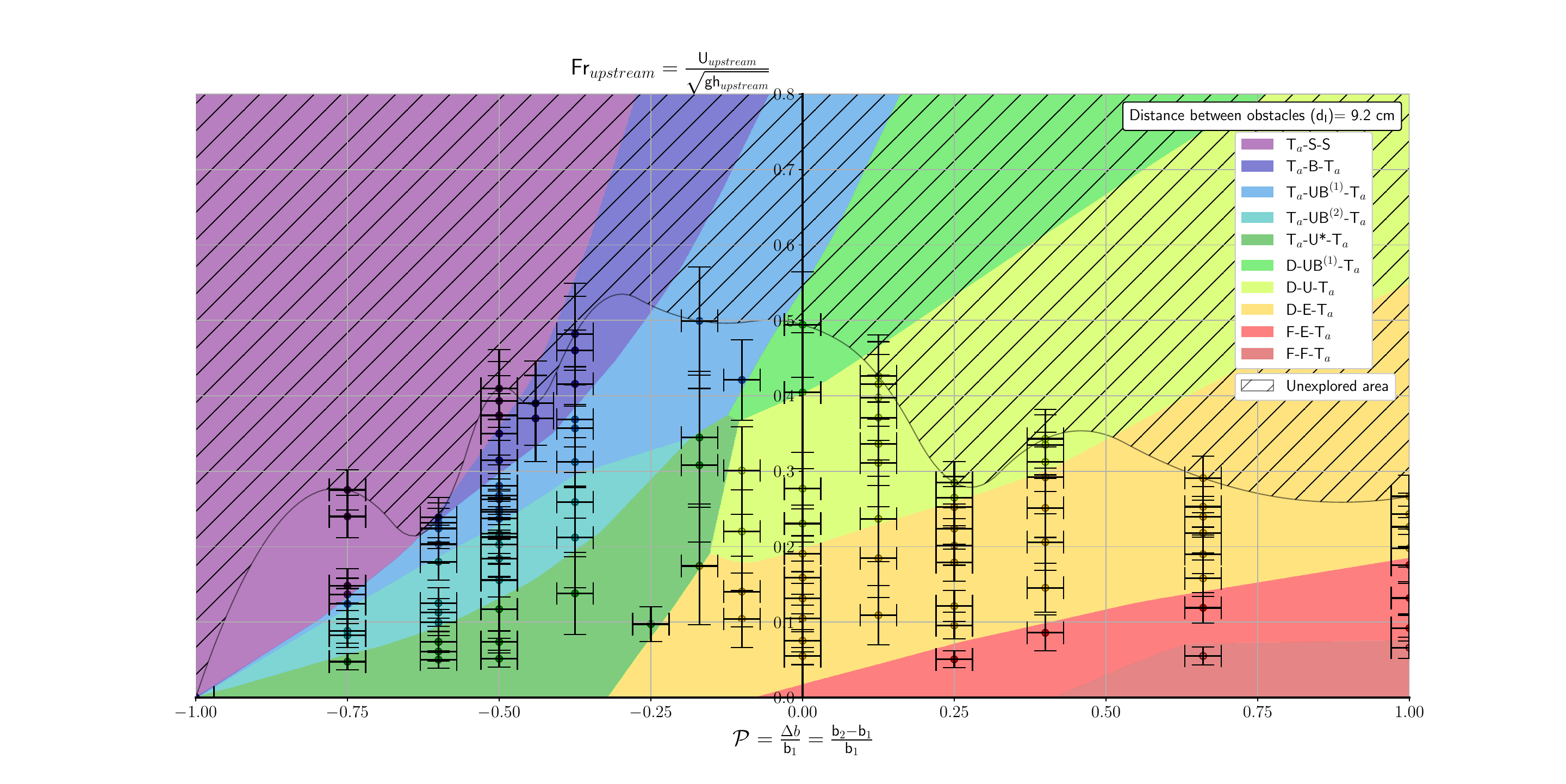}
\caption{Uncertainties of the experimental points of the phase diagram \ref{fig:diagramm}. The uncertainty $\delta \mathcal{P}$ is estimated due to the machining errors of the obstacles.
\label{incertitude diagramme de phase}}
\end{figure*}

The uncertainties used for the phase diagram \ref{fig:diagramm} are shown in the diagram \ref{incertitude diagramme de phase} for clarity. The uncertainties, for the upstream Froude number, used are statistical, (equation \ref{incert froude}). Indeed, absolute uncertainties are not used, as the resulting uncertainty intervals are overestimated. The width of the channel, W, is assumed to be known and therefore the uncertainty in the measurement of W is neglected. The uncertainty on the Pratt number is estimated to be $\delta \mathcal{P}=0.03$ because of machining uncertainties when creating obstacles.
\begin{linenomath}
\begin{equation}
        \delta Fr_{statistic}=\sqrt{\frac{\delta Q^2}{W^2gh^3}+\frac{9}{4}\frac{Q^2\delta h^2}{W^2gh^5}}
        \label{incert froude}
\end{equation}
\end{linenomath}

\subsubsection{Identification of candidate hydrodynamic regimes for the analogous black hole LASER effect.}
\begin{table*}[t]
\centering
\begin{tabular}{|c|c|c|c|c|c|c|}
\hline
  & Names of candidate regimes                                                                                             & $T_a$-$UB^{(1)}$-$T_a$                                                & $T_a$-$UB^{(2)}$-$T_a$                                                & $T_a$-$B$-$T_a$                                                       & $T_a$-$U^{*}$-$T_a$                                                   & $D$-$UB^{(1)}$-$T_a$                                                  \\ \hline
1 & \begin{tabular}[c]{@{}c@{}}Two\\ stable horizons\end{tabular}                                                          & \large\textcolor{ForestGreen}{$\surd$} &\large\textcolor{ForestGreen}{$\surd$} & \Large\textcolor{BrickRed}{$\times$}  & \large\textcolor{ForestGreen}{$\surd$} & \large\textcolor{ForestGreen}{$\surd$} \\ \hline
2 & \begin{tabular}[c]{@{}c@{}}A superluminal or\\ subliminal dispersive\\ correction\end{tabular}                         & \large\textcolor{ForestGreen}{$\surd$} & \large\textcolor{ForestGreen}{$\surd$} & \large\textcolor{ForestGreen}{$\surd$} & \large\textcolor{ForestGreen}{$\surd$} & \large\textcolor{ForestGreen}{$\surd$} \\ \hline
3 & \begin{tabular}[c]{@{}c@{}}A trapping cavity with a\\ flow regime compatible\\ with the dispersive regime\end{tabular} & \large\textcolor{ForestGreen}{$\surd$} & \large\textcolor{ForestGreen}{$\surd$} & \large\textcolor{ForestGreen}{$\surd$} & \large\textcolor{ForestGreen}{$\surd$} & \large\textcolor{ForestGreen}{$\surd$} \\ \hline
4 & \begin{tabular}[c]{@{}c@{}}Mixing of positive and \\ negative modes\end{tabular}                                       & \large\textcolor{orange}{?}      & \large\textcolor{orange}{?}      & \large\textcolor{orange}{?}      & \large\textcolor{orange}{?}      & \large\textcolor{orange}{?}      \\ \hline
5 & Avoid dissipative modes                                                                                                & \large\textcolor{orange}{?}      & \large\textcolor{orange}{?}      & \large\textcolor{orange}{?}      & \large\textcolor{orange}{?}      & \large\textcolor{orange}{?}      \\ \hline
\end{tabular}
\caption{Summary table of candidate regimes for the black hole laser effect. The table also summarises whether the regimes meet the experimental criteria for black hole lasing.}
\label{criteriaBHL}
\end{table*}

\begin{figure*}
\includegraphics[scale=0.52,width=1.05\textwidth]{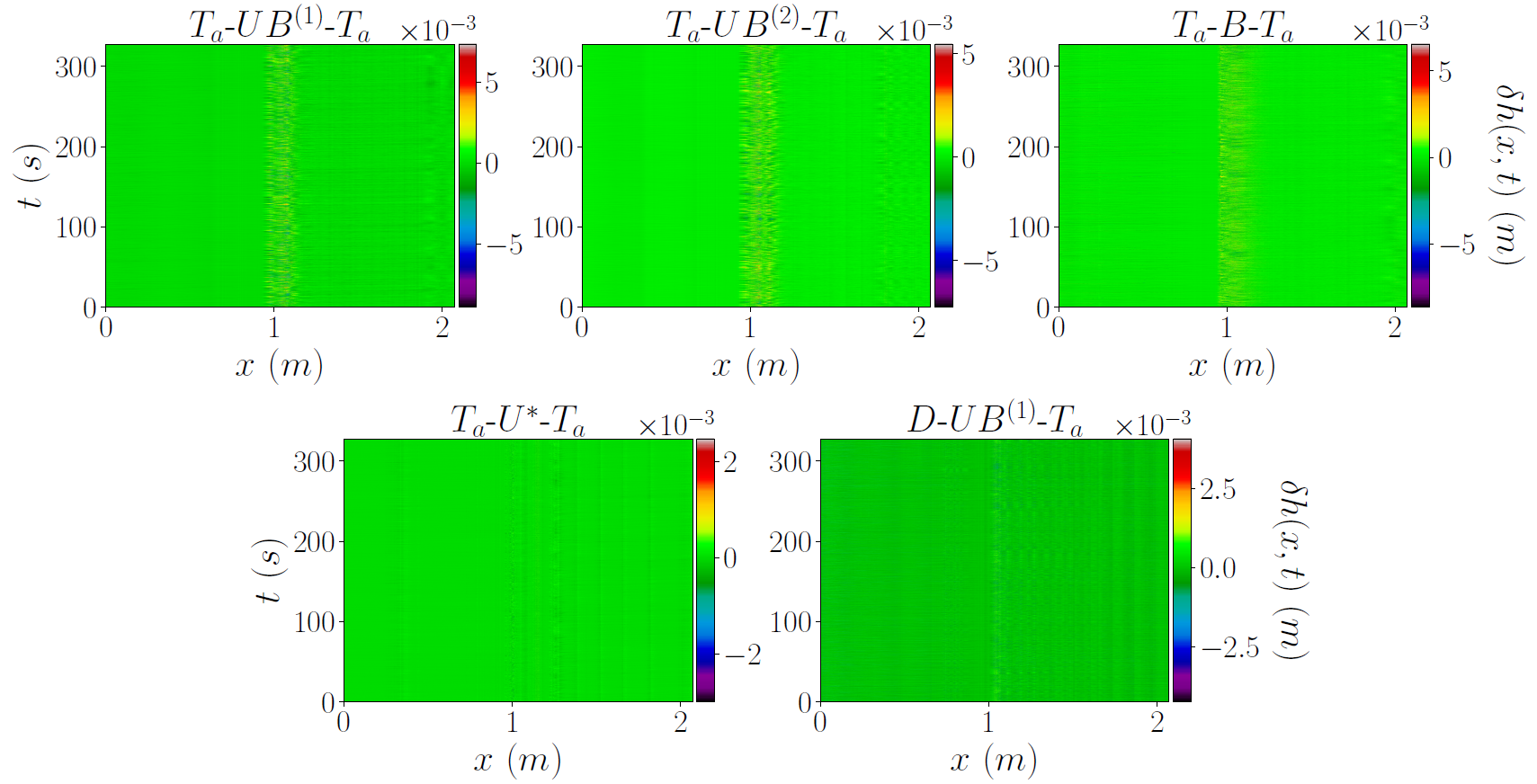}
\caption{Space-time diagram of water height fluctuations, $\delta h$, candidate hydrodynamic regimes for the black hole LASER effect. The measurement range is 2.07 m and the acquisition time is 327.6 s.
\label{regime BHL ST}}
\end{figure*}

Among these regimes, only five hydrodynamic regimes are selected for the black hole lasing effect. These regimes have been chosen because they meet the expected experimental criteria for lasing. These criteria (numbered 1 to 5) are summarised in the table \ref{criteriaBHL}. For the regimes $T_a$-$UB^{(1)}$-$T_a$, $T_a$-$UB^{(2)}$-$T_a$ and $T_a$-$B$-$T_a$, the LASER cavity could be located between both obstacles and symbolised by the cyan colour on the detection (because it is a sub-critical region with a speed superior to the Landau speed \cite{rousseaux2020classical} for the appearance of negative energy modes). The regime $T_a$-$U^{*}$-$T_a$ has two supercritical LASER cavities symbolised by the red colour on the detection. Finally, for the last regime $D$-$UB^{(1)}$-$T_a$, the cavity could be between a dispersive horizon located above the first obstacle and a black horizon (boundary from cyan to red on the detection \ref{fig:diagramm}). These regimes are candidates for the LASER effect because:
\begin{itemize}
    \item the $T_a$-$UB^{(1)}$-$T_a$ regime is closest to the prediction in general relativity \cite{CorleyJacobson}, because it is possible to have negative-energy waves (antiparticle analogue partner to Hawking radiation) in the cyan subluminal cavity;
    \item in the $T_a$-$UB^{(2)}$-$T_a$ regime, the negative-energy waves could be blocked because of capillary dispersion at small scale \cite{Rousseaux-et-al-2010};
    \item the $T_a$-$B$-$T_a$ regime has two horizons but the white horizon is unsteady as shown in the space-time diagram \ref{regime BHL ST} (hence the red cross in the table \ref{criteriaBHL});
    \item for the $T_a$-$U^{*}$-$T_a$ regime, two superluminal red cavities on both obstacles are formed but damping will kill the superluminal capillary waves \cite{Robertson-Rousseaux};
    \item and for the $D$-$UB^{(1)}$-$T_a$ regime, the undulation is trapped between a non-dispersive downstream horizon and a dispersive upstream horizon \cite{Rousseaux-et-al-2010}.
\end{itemize}
   
For the first criterion, four regimes have two stable horizons \cite{Peloquin}, with the exception of $T_a$-$B$-$T_a$. A distinction is made between horizon instability and undulation instability. In the figure \ref{regime BHL ST}, for regimes $T_a$-$UB^{(1)}$-$T_a$ and $T_a$-$UB^{(2)}$-$T_a$ the wave breaking regime appears on a stubble; for regime $T_a$-$B$-$T_a$, the wave breaking appears at the white horizon: the white horizon is unstable when the space-time diagram for this regime is scrutinized. For criteria 4 and 5, a question mark has been placed for all cases. We don't know yet whether there are positive and negative modes in the cavity and their damping, because the cavity is too small to have good resolution for the Fourier transform. A future Lasing experiment will certainly measure the presence of negative modes in cyan cavities provided the distance in-between both obstacles is sufficient to observed in the Fourier space the corresponding modes without forgetting that dissipation will dampen the modes within the cavity. The cyan colour is the insurance that surface tension will not avoid the creation of negative energy modes.

Identifying the analogue LASER effect is non-trivial and requires a detailed quantitative description, which interface detection can provide. However, the cavity is currently too small, because the obstacles are very close, to observe the modes corresponding to the Hawking effect with good spectral resolution. In addition, unanticipated phenomena such as wave breaking could perhaps be a sign of relaxation of the energy of the saturated undulation. As for the instability of the horizons that appears in the $T_a$-$B$-$T_a$ regime, the bouncing of the modes present in the cavity in the unsteady phases could explain this instability.

\subsubsection{Comparison with simulations}

\begin{figure*}
\includegraphics[scale=0.4,width=0.83\textwidth]{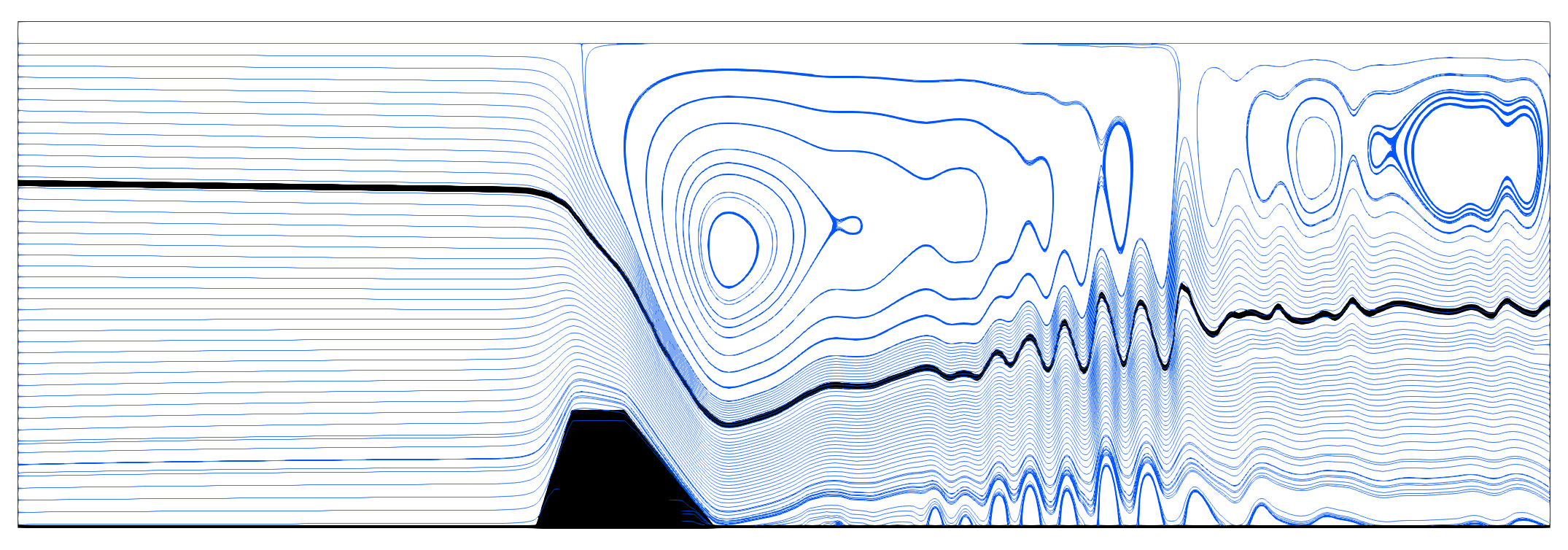}
\caption{Illustration of streamlines (blue) and water-air interface (black) provided by the Cut-Cell / Level-Set code. \label{streamlines}}
\end{figure*}

In~\cite{BDJ2012}, an original cut-cell method was developed to preserve the second order accuracy of the MAC scheme when enforcing Dirichlet boundary conditions on an obstacle of arbitrary shape. In the present paper, the cut-cell method is extended to treat two-phases incompressible flows passing over a submerged rigid obstacle. The tracking of the interface $\Gamma_t$ between the two phases (water and air) is achieved using the finite difference technique~\cite{KANG}. A rectangular domain discretized by a Cartesian grid allows to predict both water and air dynamics (see figure~\ref{streamlines}).

As we expect to get smooth interfaces, the evolution of $\Gamma_t$ is modeled 
by the level set equation 
\begin{equation}
\Gamma_t = \left\{ \phi(.,t) = 0 \right\}
\quad , \quad
\partial_t \phi + \mathbf{u} . \nabla \phi = 0 
\label{eq:levelset}
\end{equation}
where $\phi(\mathbf{x},t)$ and $\mathbf{u}(\mathbf{x},t)=(u(\mathbf{x},t),v(\mathbf{x},t))$ are the level set function and the velocity field at location $\mathbf{x}=(x,y)$ at time $t>0$.
In order to ensure that $\phi$ stay a signed distance function, i.e. $|| \nabla \phi || = 1$, the reinitialization equation 

\begin{equation}
\partial_{\tau} \phi + S(\phi_0 ) ( ||\nabla \phi|| - 1 ) =0 
\quad , \quad 
S(\phi) = \frac{ \phi }{ \sqrt{ \phi^2 + \epsilon^2 } } 
\label{eq:reinitialization}
\end{equation}
is iterated in virtual time $\tau$, where $\phi_0$ is the initial $\phi$ and $\epsilon$ is equal to the grid spacing.

Let $\delta t$ stand for the time step and $t_k = k \delta t$ discrete time values. A projection method~\cite{CHORIN} is applied on the incompressible Navier-Stokes equations to decouple velocity and pressure unknows. This leads to solve successively the prediction step: 
\begin{equation}
\frac{\mathbf{\tilde{u}}^{k+1} - \mathbf{u}^{k} }{\delta t} 
- \frac{\mu}{\rho} \triangle \mathbf{\tilde{u}}^{k+1}
= \mathbf{g} - \left( \mathbf{u}^{k} . \nabla \right) \mathbf{u}^{k}
\quad , \quad \mathbf{g} = (0,-9.81)
\label{eq:prediction}
\end{equation}
and the projection step:
\begin{eqnarray}
\frac{\mathbf{u}^{k+1} - \mathbf{\tilde{u}}^{k+1}}{\delta t}  + \frac{\nabla p^{k+1}}{\rho} = 0
\label{eq:projection} \\
\nabla . \mathbf{u}^{k+1} = 0 
\label{eq:incompressibility}
\end{eqnarray}
where $p(\mathbf{x,t})$, $\mu$ and $\rho$ are respectively the pressure, the dynamic viscosity and the density.
The variable coefficient Poisson problem
\begin{equation}
\nabla . \left( \frac{\nabla p^{k+1}}{\rho} \right) = \frac{\nabla . \mathbf{\tilde{u}}^{k+1}}{\delta t}
\label{eq:poisson}
\end{equation}
is obtained after tacking the divergence of~\eqref{eq:projection}.

A standard MAC grid is used where $p$, $\rho$ and $\phi$ exist at the cell centers, while $u$ and $v$ are located at the appropriate cell edges.
Away from both moving interface and obstacle, linear partial differential operators (divergence, gradient, Laplacian operator) and nonlinear terms are discretized on a fixed Cartesian grid by using standard second order Finite Differences approximations.

On the one hand, the Immersed Boundary Method~\cite{BDJ2012} enforce the no-slip boundary condition on the rigid obstacle. 
On the other hand, the moving interface will is tackled with the Level-Set technique~\cite{OSHER}. The level set function is evolved in time using the fifth order WENO scheme.

In the prediction step~\eqref{eq:prediction}, we simply use the sign of the level set function to determine $\rho$ as the air or water density in a sharp fashion. In particular, $\rho$ is spatially constant on either side of the interface.
The techniques presented in~\cite{LIU} for the variable coefficient Poisson equation are used to solve the equation~\eqref{eq:poisson} for the pressure. 
The resulting pressure is used to find the divergence free velocity field $\mathbf{u}^{k+1}$ defined by~\eqref{eq:projection}. However, as it was mentioned in~\cite{KANG}, one should take care to compute the derivatives of the pressure in~\eqref{eq:projection} in exactly the same way as they were computed in~\eqref{eq:poisson}.

Many direct and iterative approaches have been employed to find the solution of the linear systems~\eqref{eq:prediction} and~\eqref{eq:poisson}. For large problems, the faster solver we have found is an algebraic multigrid method (HYPRE BoomerAMG). The resolution of the linear systems has been implemented in parallel using the PETSc Fortran library~\cite{PETSC}.

A large computational domain $[-0.32~m ; 2.42~m]\times[0 ; 0.11~m]$ corresponding to the channel (figure~\ref{exp-setup}) is discretized by a $2000 \times 80$ Cartesian grid.
In addition, a constant time step $\triangle t = 1.25~10^{-4}~s$ ensures that the CFL number stay below $0.2$.

\begin{figure*}
\includegraphics[scale=0.52,width=1\textwidth]{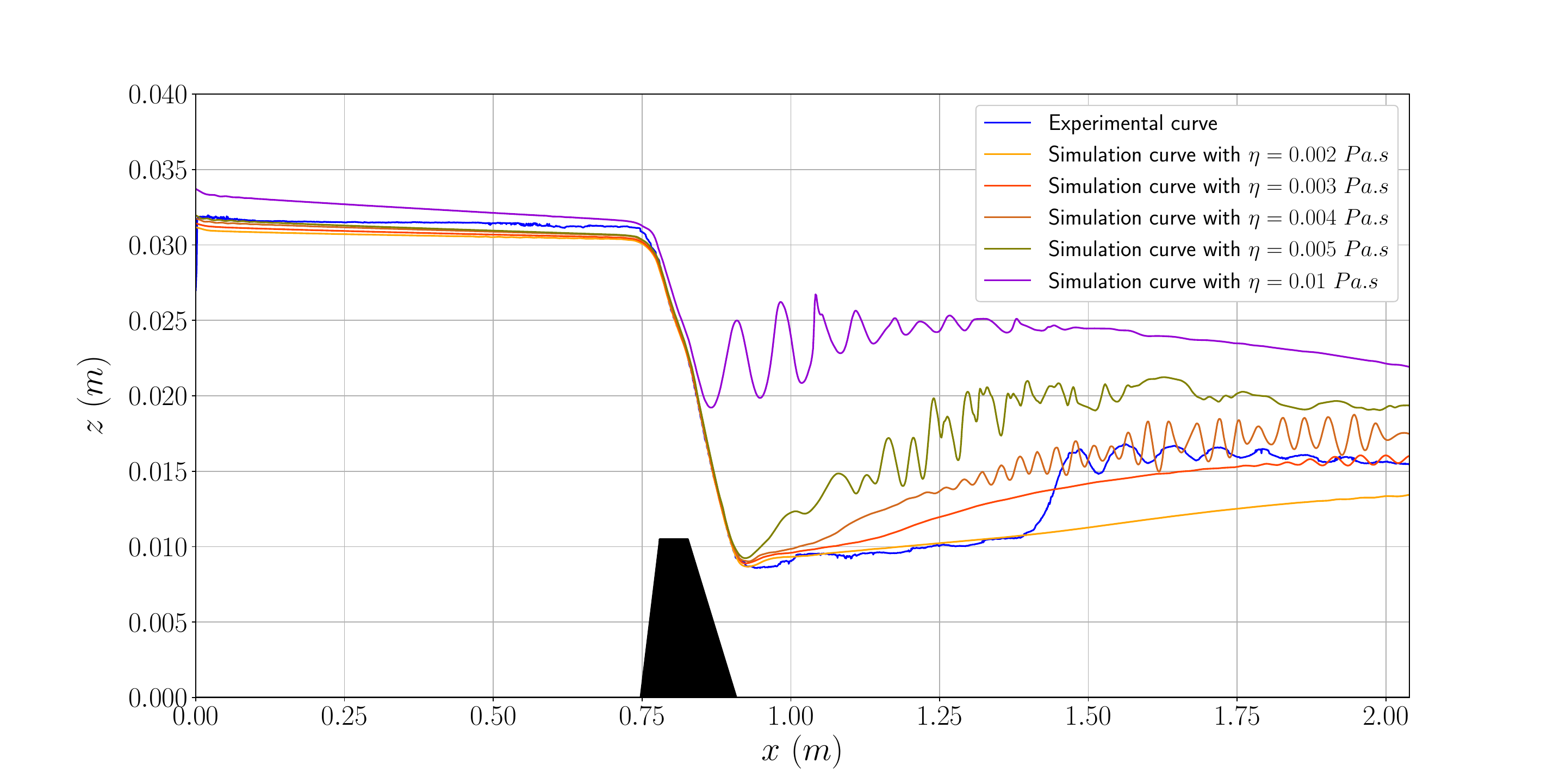}
\caption{Influence of viscosity on the water height using numerical simulations. \label{comp_visco}}
\end{figure*}

The friction of the fluid on the lateral sides of the channel is not taken into account in the two-dimensional numerical simulation. However, this induces a significant pressure drop which must be modelled. To this end, we take the dynamic viscosity $\mu$ equal to $0.005 Pa.s$ in order to fit with the experimental results. Figure~\ref{comp_visco} shows the influence  of  the viscosity $\mu$  on the  water  height  using  numerical simulations in the configuration described by figure~\ref{exemple extraction}. 

\begin{turnpage}
\begin{figure*}
\includegraphics[scale=0.5,width=1.2\textwidth]{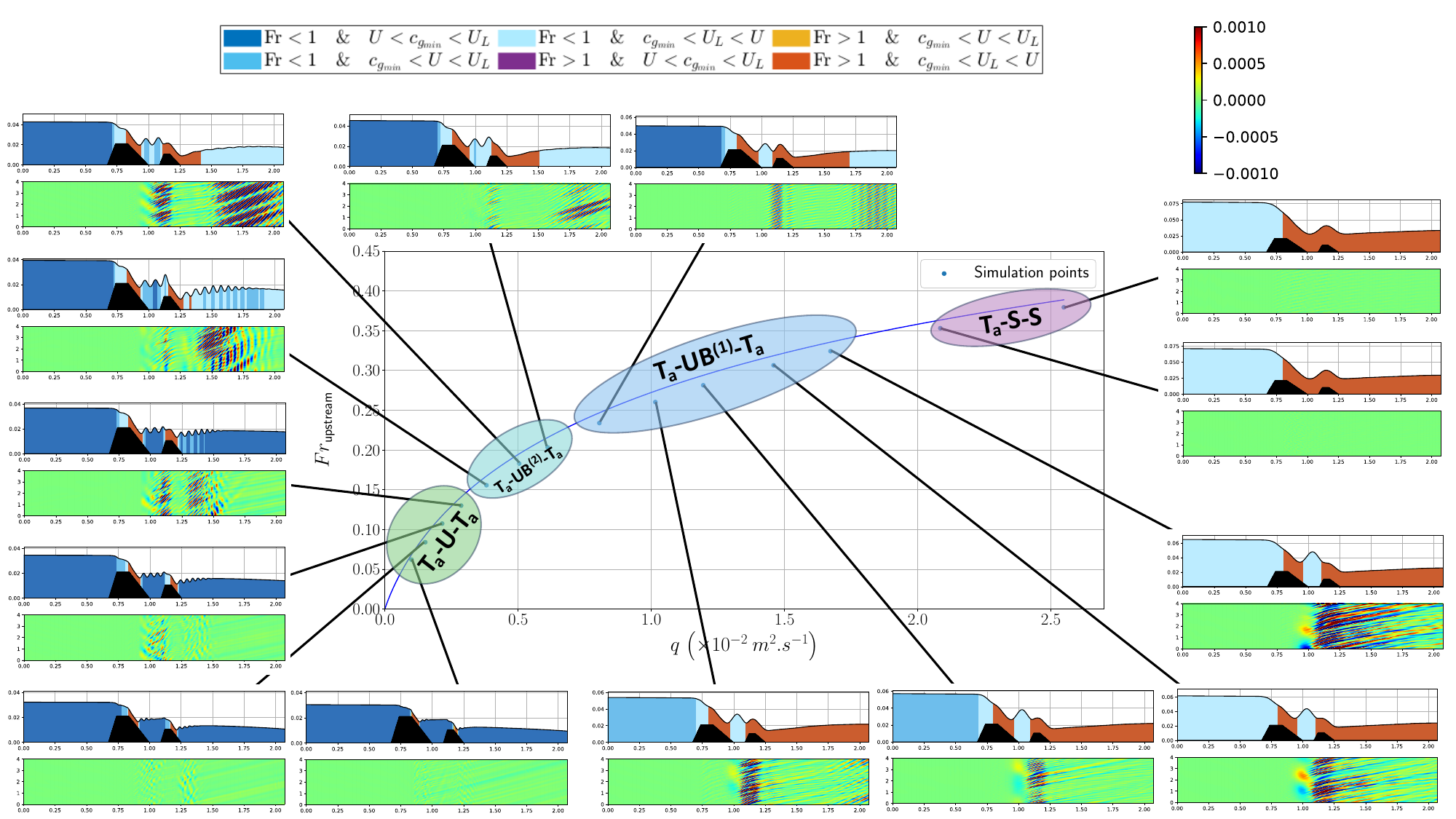}
\caption{Numerical simulations provide complementary results because, using the Cut-Cell / Level-Set code, the flow rate is not limited by the pump characteristics contrary to the experiments. We take $\mathcal{P}=-0.5$ i.e. the first obstacle is twice as large as the second. Dots corresponds to numerical simulations and the curve to Long theory \cite{Long}. For each flow rate, space-time diagram of water height fluctuations is placed under the interface extraction. \label{diagramm_simu}}
\end{figure*}
\end{turnpage}

Using flow rates between $0.001$ and $0.0255~m^2.s^{-1}$, we perform numerical simulations for $\mathcal{P}=-0.5$ i.e. the first obstacle is twice as large as the second. 
Figure~\ref{diagramm_simu} shows a good agreement with upstream Froude numbers predicted by Long theory~\cite{Long}. 
As expected, we observe $4$ differents regimes: $T_a$-$U^{*}$-$T_a$, 
$T_a$-$UB^{(2)}$-$T_a$, $T_a$-$UB^{(1)}$-$T_a$ and $T_a$-$S$-$S$. 
Again, the agreement between experimental and numerical results illustrates the efficiency and robustness of the Cut-Cell / Level-Set code.

\end{appendices}

\end{document}